\documentclass[prb,amsmath,superscriptaddress,citeautoscript,twocolumn,showpacs]{revtex4}
\usepackage{graphicx}
\usepackage{amssymb}

\newcommand{\bs}{\bar{\sigma}}

\DeclareMathOperator{\im}{Im}
\DeclareMathOperator{\re}{Re}
\newcommand{\ham}{\mathcal{H}}
\newcommand{\spinup}{\uparrow}
\newcommand{\spindown}{\downarrow}
\newcommand{\qav}[1]{\langle {#1} \rangle}
\newcommand{\qavens}[1]{\mathbf{\ll} {#1} \mathbf{\gg}}

\begin{document}
\title{On the applicability of the equations-of-motion technique for
quantum dots}

\author{Vyacheslavs Kashcheyevs}
\email[e-mail: ]{slava@latnet.lv}
\affiliation{School of Physics and Astronomy,
Raymond and Beverly Sackler Faculty of Exact Sciences, \\
Tel Aviv University, Tel Aviv 69978, Israel}

\author{Amnon Aharony}
\affiliation{School of Physics and Astronomy,
Raymond and Beverly Sackler Faculty of Exact Sciences, \\
Tel Aviv University, Tel Aviv 69978, Israel}

\affiliation{Department of Physics, Ben Gurion University,
Beer Sheva 84105, Israel}

\author{Ora Entin-Wohlman}
\affiliation{School of Physics and Astronomy,
Raymond and Beverly Sackler Faculty of Exact Sciences, \\
Tel Aviv University, Tel Aviv 69978, Israel}

\affiliation{Department of Physics, Ben Gurion University, Beer
Sheva 84105, Israel}

\affiliation{Albert Einstein Minerva Center for Theoretical
Physics \\ at the Weizmann Institute of Science,  Rehovot 76100,
Israel}


\begin{abstract}
The  equations-of-motion (EOM) hierarchy satisfied by the Green
functions of a quantum dot embedded in an external mesoscopic
network is considered within a high-order decoupling approximation
scheme. Exact analytic solutions of the resulting coupled integral
equations are presented in several limits. In particular, it is
found that at the particle-hole symmetric point the EOM Green
function is temperature-independent due to a discontinuous change
in the imaginary part of the interacting self-energy.  However,
this imaginary part obeys the Fermi liquid unitarity requirement
away from this special point, at zero temperature. Results for the
occupation numbers, the density of states and the local spin
susceptibility are compared with exact Fermi liquid relations and
the Bethe \emph{ansatz} solution. The approximation is found to be
very accurate far from the Kondo regime. In contrast, the
description of the Kondo effect is valid on a qualitative level
only. In particular, we find that the Friedel sum rule is
considerably violated, up to $30 \%$, and the spin susceptibility
is underestimated. We show that the widely-used simplified version
of the EOM method, which does not account fully for the
correlations on the network, fails to produce the Kondo
correlations even qualitatively.
\end{abstract}

\maketitle

\section{Introduction}
Quantum dots embedded in mesoscopic structures are currently of
great experimental and theoretical interest, because such systems
allow for detailed and controlled studies of the effects of
electronic correlations\cite{Kouwenhoven97}. The theoretical
description of these systems is usually based on the Anderson
model\cite{Anderson61,NgLee88GlazmanRaikh88}, in which the
electronic correlations are confined to  few impurities that
represent the quantum dots. Although a rich variety of techniques
has been developed over the years to treat the  Anderson model
\cite{Hewson93},  their applications to the quantum dot systems
are not straightforward. In such systems one would like to be able
to study dynamical properties (e.g.,  transport) as function of
the impurity characteristics (which can be tuned experimentally)
over a wide range of parameters. This is not easily accomplished
by the Bethe \emph{ansatz} solution, for example.

The theoretical difficulty can be pinned down to the ability to
derive a reliable, easy-to-handle, expression for the
single-electron Green function on the quantum dot. Because the
electronic interactions in the Anderson model take place solely on
the dot, this Green function can be shown to determine, under
certain conditions, the charge or the spin transmission through
the quantum dot,  the charge accumulated on it, etc
\cite{NgLee88GlazmanRaikh88,Meir91ANDGF}. Consequently, much
effort has been devoted to finding faithful analytic
approximations for this object. Alternative treatments rely on
numerical techniques, such as quantum Monte
Carlo\cite{Hirsch86Silver90}, or the numerical renormalization
group (NRG) method \cite{NrgMethods}.   NRG in particular is
considered to be capable of providing accurate estimates of the
Green function over a wide parameter range, although at the cost
of running iterative diagonalizations for each parameter set, and
a limited resolution at high energies and high magnetic fields.

A ubiquitous method to derive an analytical expression for the
Green functions is to use the equations-of-motion
(EOM)\cite{Zubarev60}. In the case of the single-impurity Anderson
model, the EOM of the (single-particle) impurity Green function
gives rise to an infinite hierarchy of EOM of higher-order Green
functions. A well-known approximation procedure is then to
truncate this hierarchy, thus producing certain thermal averages
representing various correlation functions. The latter need to be
found self-consistently from the resulting closed set of
equations. The level at which the EOM are truncated is chosen such
that most of the interaction effects are captured.
\cite{Theumann69, Appelbaum69both,Lacroix81,Lacroix82} This scheme
has been applied to the original Anderson model  a long time ago,
\cite{Theumann69,
Appelbaum69both,MamadaTakano70,Lacroix81,GeeSweePoo75both}
yielding approximate expressions for the resistivity,
\cite{Theumann69,Appelbaum69both} the spin susceptibility,
\cite{Appelbaum69both,MamadaTakano70} and the magneto transport
\cite{GeeSweePoo75both} of dilute magnetic alloys. For
temperatures above the Kondo temperature, $T_K$, these results
agree with perturbation theory calculations. \cite{Hewson93}
Although several limitations of these self-consistent approaches
are known (such as underestimation of the Kondo temperature or the
absence of $(T/T_{K})^2$ terms in the low-temperature expansion of
the results), they still can form a basis for a qualitative
analytic treatment of the Kondo effect.

The first application of the EOM technique to a quantum dot system
has been undertaken by Meir, Wingreen and Lee\cite{Meir91ANDGF}
(MWL). Neglecting certain correlation functions, they have
obtained a closed analytic expression for the dot Green function.
Several subsequent studies have employed their scheme  to describe
various effects of the Kondo correlations in quantum dots in
different settings, e.g., ac response of a biased quantum dot,
\cite{Ng96meircite} non-equilibrium Andreev tunneling,
\cite{FazioRaimondi98SunGuoLin01} and coupling to magnetic leads.
\cite{Zhang02Sergueev02} As we show below, the approximation of
MWL fails in various aspects. Recent attempts to improve on that
solution turned out to be not completely satisfactory: either
requiring further approximations on top of the self-consistent
truncation, \cite{EntinAharonyMeir04} or leading to iterative
numerical solutions\cite{Bulka01,Luo02,Bulka03}. It is therefore
of interest to examine whether an \emph{exact solution} of the
self-consistency equations of the EOM method will improve on the
previous results, and will be capable of producing a reliable
approximation for the Green function that can be used in the
analysis of quantum dot systems.

In this paper we consider a quantum dot embedded in a general
complex network, and obtain and solve the truncated EOM for its
Green function. This EOM contains  all the next order correlation
functions, which we decouple and calculate self-consistently. We
first derive (in Sec. \ref{sec:def}) an integral equation for the
Green function, allowing for arbitrary values of the on-site
Coulomb interaction, $U$. We then analyze (in Sec.
\ref{sec:finiteU}) the properties of this EOM solution. We find
that in the case of a particle-hole symmetric Hamiltonian the
resulting Green function is \emph{temperature-independent} (a
point which is overlooked in previous treatments of quantum dots,
see for example Ref.~~\onlinecite {Meir91ANDGF}). However, we show
that this is a singular point in the parameter space. We also
investigate the zero-temperature limit of the EOM solution, and
show that it fails to satisfy the Friedel sum-rule. We then turn
to the infinite $U$ case (in Sec. \ref{sec:Uinf}), and derive an
exact analytical solution to the integral equation for the Green
function. In Sec. \ref{sec:physUinf} we use this solution to
obtain the total occupation number on the dot, and compare it with
the exact solution supplied by the Bethe {\em ansatz}. This
comparison shows that the EOM solution is faithful outside the
Kondo regime, but fails in the regime where Kondo correlations
play a dominant role. In particular, the Friedel sum-rule is
violated by $\sim 30\%$, invalidating the assumptions made in
previous studies \cite{Lacroix81,Bulka01,Bulka03}. We then derive
the local spin-susceptibility on the dot and demonstrate that the
full self-consistent solution of the EOM, as used in this work, is
required in order to obtain quantitatively correct results. We
also examine the local density of states at the Fermi level, and
find that it shows the expected universal behavior as function of
$T/T_{K}$, though the EOM Kondo temperature  lacks a factor of two
in the exponential dependence on the single-energy level on the
dot.  Finally, we examine the EOM technique from another point of
view (Sec. \ref{PT}): We expand the Green function derived from
the EOM to second order in the dot-network coupling, and compare
the results with those obtained from a straightforward
perturbation theory \cite{Sivan96,Gefen04}. This comparison shows
again the necessity to include in the EOM solution the full
self-consistent calculation of all the correlations. A short
summary in Sec.~\ref{sec:conclusions} concludes the paper.

\section{The Green function on the dot\label{sec:def}}
As mentioned above, several properties of the Anderson model can
be expressed in terms of the Green function on the dot. Here we
examine the determination of this function using the EOM method.
Our discussion is limited to a single interacting impurity
embedded in a general non-interacting network, for which the
Hamiltonian can be written in the form
\begin{eqnarray}
  \ham=\ham_{\text{dot}}+\ham_{\text{net}}+\ham_{\text{net-dot}} \, .\label{HAM}
\end{eqnarray}
Here, the dot Hamiltonian is given by
\begin{eqnarray}
  \ham_{\text{dot}}=\sum_{\sigma} (\epsilon_0+ \sigma  h )
  n_{d\sigma} + U n_{d\spinup}  n_{d\spindown}\, , \;\;
  n_{d\sigma}=d^{\dagger}_{\sigma}d^{}_{\sigma}\, ,
\end{eqnarray}
where $d^{\dagger}_{\sigma}$ ($d_{\sigma}$) is the creation
(annihilation) operator of an electron of spin index $\sigma=\pm
1/2$ on the dot, $\epsilon_0$ is the single-particle energy there,
$h$ is the Zeeman splitting, and $U$ denotes the Coulomb repulsion
energy. The  non-interacting network is described by the
tight-binding Hamiltonian
\begin{eqnarray}
  \ham_{\text{net}}=\sum_{n\sigma} \varepsilon_{n\sigma}
  a^{\dagger}_{n\sigma} a^{}_{n\sigma} - \sum_{mn\sigma} J^{}_{mn}
  a^{\dagger}_{m\sigma} a^{}_{n\sigma} \, ,
\end{eqnarray}
where $a^{\dagger}_{n\sigma}$ ($a^{}_{n\sigma}$) is the creation
(annihilation) operator of an electron of spin index $\sigma$ on
the $n$th site on the network, whose on-site energy is
$\varepsilon_{n\sigma} $, and $J^{}_{mn}=J^{\ast}_{nm}$ are the
hopping amplitudes on the network. Finally, the coupling between
the dot and the network is given by
\begin{eqnarray}
  \ham_{\text{net-dot}}= -\sum_{n\sigma} J^{}_{n\sigma}
  d^{\dagger}_{\sigma} a^{}_{n\sigma} +\text{H.c.}
\end{eqnarray}
We have allowed for spin-dependent on-site energies on the
network, as well as spin-dependent hopping amplitudes between the
dot and the network. In this way, our model includes also the case
of spin-polarized leads connected to a quantum dot (see for
example, Refs.~~\onlinecite{Meir91ANDGF} and
~\onlinecite{Bulka03}). The entire system is assumed to be at
equilibrium with a reservoir held at temperature $T$ (in energy
units) and chemical potential $\mu=0$.

Adopting the notations of Ref.~~\onlinecite{Zubarev60}, we
write a general Green function in the form
\begin{eqnarray}
  &&\qavens{A; B}_{\omega\pm i \eta}\nonumber\\
  && \equiv \mp i \int_{-\infty}^{+\infty} \Theta(\pm t) \qav{[A(t);
  B]_{+}} e^{i (\omega \pm i\eta) t} dt \, , \label{GREEN}
\end{eqnarray}
where $A$ and $B$ are operators, $\Theta$ is the Heaviside
function, and $\eta\rightarrow 0^{+}$. The Green function on the
dot is then
\begin{eqnarray}
  G_{\sigma}(z)\equiv \qavens{d^{}_{\sigma};d^{\dagger}_{\sigma}}_z
  \, , \;\; z\equiv \omega\pm i\eta \, .
\end{eqnarray}
In conjunction with the definition (\ref{GREEN}), a thermal
average, $\qav{BA}$, is related to the corresponding
Green function by
\begin{eqnarray}\label{eq:FDtheorem}
  \qav{BA}=i \oint_C\frac{dz}{2\pi} f(z) \qavens{A; B}_z \, , \;\;
  f(z)\equiv \frac{1}{1+e^{z/T}}\, ,
\end{eqnarray}
where the contour $C$ runs clockwise around the real axis.

The EOM for the dot Green function is given in Appendix
\ref{sec:appendixA} [see Eq. (\ref{GdExact})]. As is shown there,
that equation includes a higher-order Green function, whose EOM
gives rise to additional Green functions. The resulting infinite
hierarchy of EOM is then truncated according to a scheme proposed
originally by Mattis \cite{foot1} and subsequently used in
Refs.~~\onlinecite{Theumann69,Appelbaum69both,Lacroix81,
MamadaTakano70,GeeSweePoo75both,
Bulka01,Luo02,Bulka03,EntinAharonyMeir04,Luo99}: Each Green
function of the type $\qavens{ A^{\dagger} B C ; d^{\dagger}}$ in
the EOM hierarchy is replaced by
\begin{eqnarray} \qavens{
  A^{\dagger} B C ; d^{\dagger}} \Rightarrow \qav{A^{\dagger} B}
  \qavens{ C; d^{\dagger}} - \qav{A^{\dagger} C} \qavens{ B;
  d^{\dagger}} \label{trunc}
\end{eqnarray}
\emph{if at least two of
the operators $A$, $B$ and $C$ are network operators
$a_{n\sigma}$}. Explicitly, the Green functions which are
decoupled are $\qavens{a^{}_{n\sigma} d^{\dagger}_{\bar{\sigma}}
a^{}_{m\bar{\sigma}} ;
 d^{\dagger}_\sigma}$,
$\qavens{a^\dagger_{n \bar{\sigma}} d^{}_{\bar{\sigma}}
a^{}_{m\sigma} ; d^{\dagger}_\sigma}$, and $\qavens{a^{\dagger}_{n
{\bar{\sigma}}} a^{}_{m{\bar{\sigma}}} d^{}_\sigma ;
d^{\dagger}_\sigma}$ (with $\bar{\sigma}=-\sigma$).\cite{foot2}
Upon calculating the averages $\qav{a^\dagger_{m \sigma}
a^{}_{m' \sigma}}$ and $ \qav{d^\dagger_{\sigma}
a^{}_{m\sigma}}$ using Eq. (\ref{eq:FDtheorem}), the set of
EOM is closed, and can be therefore solved. The details of this
calculation are presented in Appendix \ref{sec:appendixA}. In
particular, the resulting equation determining the dot Green
function is
\begin{widetext}
\begin{align}
  G_{\sigma}(z) = \frac{u(z)-\qav{n_{d\bar{\sigma}}}
  -P_{\bar{\sigma}}(z_1) -P_{\bar{\sigma}}(z_2)} {u(z) \left
  [z-\epsilon_0 - \sigma h -\Sigma_{\sigma}(z)\right]+\left [
  P_{\bar{\sigma}}(z_1) +P_{\bar{\sigma}}(z_2)\right]
  \Sigma_{\sigma}(z) -Q_{\bar{\sigma}}(z_1) +Q_{\bar{\sigma}}(z_2)}
  \, , \label{eq:GfiniteU}
\end{align}
where
\begin{eqnarray}
  u(z)\equiv U^{-1} \left [ U-z+\epsilon_0+ \sigma
  h+\Sigma_{\sigma}(z)+\Sigma_{\bar{\sigma}}(z_1)
  -\Sigma_{\bar{\sigma}}(z_2) \right] \, ,\label{u}
\end{eqnarray}
\end{widetext}
and $z_1\equiv z-2 \sigma h$, $z_2 \equiv -z + 2 \epsilon_0+U$.
The functions $P$ and $Q$ are given in terms of the
non-interacting self-energy on the dot, $\Sigma_{\sigma}(z)$,
brought about by its coupling to the network [namely, the
self-energy of the non-interacting dot, see Eq. (\ref{Sigmadef})],
and the dot Green function itself [see Eqs. (\ref{EXPLICIT})],
\begin{align}
  & P_\sigma(z)   = \mathfrak{F}_{\sigma z} \bigl[G] \; , \,
  Q_\sigma(z)   = \mathfrak{F}_{\sigma z} \bigl[1+\Sigma \, G] \; ,
  \label{eq:PQselfcon}
\end{align}
where the notation $\mathfrak{F}_{\sigma z}[g]$ stands for
\begin{align}
& \mathfrak{F}_{\sigma z}[g]\equiv \frac{i}{2 \pi}\oint_C f(w)
g_\sigma(w) \frac{\Sigma_{\sigma}(w)-\Sigma_{\sigma}(z)}{z-w} d w
\, . \label{eq:defThermal}
\end{align}
Equation (\ref{eq:GfiniteU}) generalizes the result of
Ref.~~\onlinecite{Lacroix81} (see also
Refs.~~\onlinecite{Bulka01,Luo02,Bulka03,EntinAharonyMeir04}) for
the case in which the interaction on the dot is finite, and the
entire system is subject to an external magnetic field. Our
generalization also corrects a few details in Lacroix's earlier
treatment of finite $U$. \cite{Lacroix82}

\section{Properties of the EOM approximation at finite $U$}
\label{sec:finiteU}
As is evident from Eq. (\ref{eq:GfiniteU}),
the solution of the dot Green function within the EOM scheme
cannot be easily obtained over the entire parameter range.
However, there are certain limiting cases in which this Green
function can be analyzed analytically. We examine those in the
subsequent subsections.

\subsection{Particle-hole symmetry\label{sec:phsym}}
Upon replacing the particle operators by the hole ones,
$\tilde{d}_{\sigma}^{\dagger} \equiv d^{}_{\sigma},
\tilde{a}_{n\sigma}^{\dagger} \equiv a^{}_{n\sigma}$, the Anderson
Hamiltonian (\ref{HAM}) attains its original structure, with
\begin{eqnarray}
&&\tilde{\epsilon}_0 + \sigma \tilde{h} =-\epsilon_0-\sigma h-U \, , \;\;
\tilde{U}=U \; , \nonumber\\
&&\tilde{J}^{}_{n\sigma} =-J^{\ast}_{n\sigma} \, , \;\;
\tilde{\varepsilon}_{n\sigma} = -\varepsilon_{n\sigma}\, ,\;\;
\tilde{J}_{nm} =-J_{mn} \, . \label{eq:phrules}
\end{eqnarray}
(Hole quantities are denoted by a tilde.) The dot Green function
in terms of the hole operators is then related to the particle
Green function by
\begin{align}
   \tilde{G}_{\sigma}(z)\equiv
   \qavens{\tilde{d}^{}_{\sigma};\tilde{d}^{\dagger}_{\sigma}}_z
  =-G_{\sigma}(-z) \,  . \label{eq:holeG}
\end{align}
One may check that this equivalence holds by introducing the
definitions (\ref{eq:phrules}) into Eq.~\eqref{eq:GfiniteU}. Since
$\tilde{\Sigma}_{\sigma}(z)=-\Sigma_{\sigma}(-z)$ and
$\tilde{u}(z)=1-u(-z)$, one  finds that [see Eqs. (\ref{EXPLICIT})
and (\ref{eq:integralexample})]
$\tilde{P}_{\sigma}(z)=-{P}_{\sigma}(-z)$ and
$\tilde{Q}_{\sigma}(z)={Q}_{\sigma}(-z)-\Sigma_{\sigma}(-z)$,
re-confirming Eq. (\ref{eq:holeG}).

From now on we shall assume that
$\Sigma_{\sigma}(z)=-\Sigma_{\sigma}(-z)$. This relation  is
realized, for example, when the network to which the dot is
coupled has a wide band spectrum, with the Fermi level at the
middle.  We next discuss  the particular point where $2
\epsilon_0+U=0$ and $h=0$. At this point, the Anderson Hamiltonian
becomes \emph{particle-hole symmetric}.  Then
$G_{\sigma}(z)=-G_{\sigma}(-z)$, $P_{\sigma}(z)+P_{\sigma}(-z)=0$,
$Q_{\sigma}(z)-Q_{\sigma}(-z)=\Sigma_{\sigma}(z)$, and
$\qav{n_{d\sigma}}=1/2$. As a result,  Eq. (\ref{eq:GfiniteU})
becomes
\begin{eqnarray}\label{eq:symsol}
\left[ G_{\sigma}(z) \right ]^{-1}&=&
z-\Sigma_{\sigma}(z)\nonumber\\
&-&\frac{U^2} {4 \left[z-\Sigma_{\sigma}(z)-2
\Sigma_{\bar{\sigma}}(z) \right ]} \,\  .
\end{eqnarray}
Namely, at the particle-hole symmetry point of the Anderson model,
the EOM results in a \emph{temperature-independent} dot Green
function! This implies that the EOM technique at the particle-hole
symmetric point \emph{cannot} produce the Kondo singularity. This
property of the EOM scheme has been reported a long time ago,
\cite{Appelbaum69both}${}^{\text{a},}{}$\cite{Dworin67both,Oguchi70}{}
but was ignored in more modern uses of it, \cite{Meir91ANDGF}
which are designed to study the Kondo peak in the density of
states.

The failure of the EOM method to describe faithfully the Anderson
model at its symmetric point, where the Fermi level lies
\emph{exactly} between the states of single and double
occupancies, is a very severe drawback of this method. An
important question is whether this point is singular, or is there
a continuous domain in which the EOM method fails totally. We
return to this problem in the next subsection.

\subsection{Zero-temperature relations\label{sec:zerotemp}}
The zero-temperature limit is of special importance since  the
Green function \emph{at the Fermi energy} at $T=0$ satisfies the
Fermi-liquid relations \cite{Hewson93,Langreth66}
\begin{align}
  \im [G^{+}_{\sigma}(0)]^{-1} & =   \Gamma_{\sigma} \, ,
  \label{eq:unitarity} \\
  \re [G^{+}_{\sigma}(0)]^{-1} &=  - \Gamma_{\sigma} \cot ( \pi
  \tilde{n}_{d\sigma} ) \, . \label{FRIEDEL}
\end{align}
Here and below we use  $A^{\pm}(\omega)\equiv \lim_{\eta\to\pm0}
A(\omega+i \eta)$, so that
$G^{+}_{\sigma}(\omega)=[G^{-}_{\sigma}(\omega)]^{\ast}$ is the
usual retarded Green function. In Eqs. (\ref{eq:unitarity}) and
(\ref{FRIEDEL}), $\Gamma_{\sigma}$ is the level broadening,
\begin{eqnarray}
  \Gamma_{\sigma}\equiv -\im \Sigma_{\sigma}^{+}(0) \, .
\end{eqnarray}
The first relation, Eq. (\ref{eq:unitarity}), implies
\cite{Langreth66} number conservation, and therefore is sometimes
referred to as the `unitarity' condition. %\cite{schiller}
The second one, Eq. (\ref{FRIEDEL}), is the Friedel sum-rule, in
which $\tilde{n}_{d\sigma}$ is the total number of spin $\sigma$
electrons introduced by the quantum dot, \cite{Langreth66}
\begin{align}
  \tilde{n}_{d\sigma} & = - \frac{1}{\pi}  \im \int f(\omega) \left
  [ 1- \frac{ \partial \Sigma^{+}_\sigma(\omega)}{\partial \omega}
  \right ] G_{\sigma}^{+}(\omega) \, d \omega \, .\label{ntild}
\end{align}
When the self-energy $\Sigma_{\sigma}$ does not depend on the
energy, $\tilde{n}_{d\sigma}$ coincides with the single-spin
occupation number on the dot, $\qav{n_{d\sigma}}$.

It is evident that the EOM solution for the Green function
\emph{at the particle-hole symmetric point}, Eq.
(\ref{eq:symsol}), \emph{ violates} the unitarity condition
(\ref{eq:unitarity}). On the Fermi level, the particle-hole
symmetric Green function is
\begin{eqnarray}
  [G^{+}_{\sigma}(0)]^{-1}=i\Gamma_{\sigma}+ i\frac{U^{2}}{4\bigl
  (\Gamma_{\sigma}+2\Gamma_{\bar{\sigma}}\bigr )}\, ,\label{SYM0}
\end{eqnarray}
and therefore the imaginary part of $[G^{+}_{\sigma}(0)]^{-1}$ is
\emph{not} determined solely by the non-interacting self-energy,
(as implied by the unitarity condition), but  has also  a
contribution coming from the interaction. This is contrary to the
result of Lacroix, \cite{Lacroix82} whose EOM differs from our Eq.
(\ref{eq:GfiniteU}) in several places. On the other hand, the
Friedel sum-rule is satisfied by the Green function
(\ref{eq:symsol}), which yields $\tilde{n}_{d\sigma}=1/2$. This
follows from Eq. (\ref{ntild}): The imaginary parts of both
$G_{\sigma}(\omega )$ and $\Sigma_{\sigma}(\omega )$ are even in
$\omega$, while (at the symmetric point) the real parts are odd in
it. However, $G_{\sigma}(\omega )\simeq \omega^{-1}$ at large
frequencies, whereas $\partial\Sigma_{\sigma}(\omega
)/\partial\omega \simeq\omega^{-2}$. As a result, the second term
in the square brackets of Eq. (\ref{ntild}) does not contribute.
With $\tilde{n}_{d\sigma}=1/2$, the Friedel sum-rule gives
$\re[G^{+}_{\sigma}(0)]^{-1}=0$, which is fulfilled by Eq.
(\ref{SYM0}).

It is rather intricate to study the full EOM solution, Eq.
(\ref{eq:GfiniteU}), at $T=0$, even on the Fermi level. However,
there are  cases in which this can be accomplished without
constructing the full solution. The investigation of these cases
will also allow us to examine the behavior of $G_{\sigma} (0)$ as
the particle-hole symmetric point is approached. To this end we
note that at $T=0$ the functions $P(\omega )$ and $Q(\omega )$
acquire logarithmic singularities as $\omega \to 0$,
\begin{align}
  P_{\sigma}^{\pm}(\omega) & \sim - \frac{1}{\pi}
  \Gamma_{\sigma} G^{\mp}_{\sigma} (0) \ln |\omega|+\mathcal{O}(1)
  \,  , \nonumber\\
  Q_{\sigma}^{\pm}(\omega) & \sim -\frac{1}{\pi} \Gamma_{\sigma}
  \bigl[1+\Sigma^{\mp}_{\sigma}(0) G^{\mp}_{\sigma} (0)\bigr] \ln
  |\omega|+\mathcal{O}(1)\, .\label{LOG}
\end{align}
Therefore, we may examine special points at which the functions
$P$ and $Q$ are divergent,  keeping only the divergent terms in
Eq. (\ref{eq:GfiniteU}). Then, that equation reduces to an
algebraic one, which can be easily solved to yield $G_{\sigma}$ at
those special points.

Let us first consider the case in which the Zeeman field $h$ on
the dot vanishes, but $2\epsilon_{0}+U\neq 0$. Using Eqs.
(\ref{LOG}) in Eq. (\ref{eq:GfiniteU}) yields
\begin{eqnarray}
  [G^{+}_{\sigma}(0)]^{-1}+ \Sigma_{\sigma}^{+}(0)=
  [G_{\bar{\sigma}}^{-}(0)]^{-1} +\Sigma_{\bar{\sigma}}^{-}(0) \, .
  \label{eq:hT0}
\end{eqnarray}
By writing the Green function in the general form
\begin{eqnarray}
  [G^{}_{\sigma}(z)]^{-1}=z-\epsilon_{0}-\sigma
  h-\Sigma^{}_{\sigma}(z)-\Sigma^{\text{int}}_{\sigma} (z)
  \, , \label{GENE}
\end{eqnarray}
in which $\Sigma^{\text{int}}$ is the self-energy due to the
interaction, Eq.~(\ref{eq:hT0}) takes the form
\begin{eqnarray}
\Sigma^{\text{int}\,+}_{\sigma}(0)-\Sigma^{\text{int}\,-}_{\bar{\sigma}}(0)
=0 \, .
\end{eqnarray}
When the network is not spin-polarized, the spin indices $\sigma$
and $\bar{\sigma}$ are indistinguishable. Then Eq.~(\ref{eq:hT0})
implies that $\im\Sigma^{\text{int}}(0)$ vanishes, namely the
unitarity condition is satisfied. (In the more general case of
possibly ferromagnetic leads, it is only the imaginary part of
$[G_{\sigma}^{+}(0)]^{-1}-[G_{\bar{\sigma}}^{-}(0)]^{-1}$ which is
determined by the non-interacting self-energy alone.) Had we now
sent $2\epsilon_{0}+U$ to zero, we would have found that the EOM
result at the particle-hole symmetric point \emph{does} satisfy
the unitarity condition, in contradiction to our finding,
Eq.~(\ref{SYM0}), above. We thus conclude that the failure of the
EOM to obey the Fermi-liquid relation (\ref{eq:unitarity}) at the
symmetric point is \emph{confined} to the symmetric point alone,
namely, the imaginary part of $\Sigma^{\text{int}}$ on the Fermi
level has a discontinuity.

Next we consider the case where $2\epsilon_{0}+U= 0$, but $h\neq
0$. Using Eqs. (\ref{LOG}) in Eq. (\ref{eq:GfiniteU}) we now find
the relation
\begin{align}\label{eq:h2eU0}
  [G^{+}_{\sigma}(0)]^{-1}+\Sigma^{+}_{\sigma}(0)= -\bigl \{
  [G^{+}_{\bar{\sigma}}(0)]^{-1} +\Sigma^{+}_{\bar{\sigma}}(0) \bigr
  \} \, .
\end{align}
Inserting here expression (\ref{GENE}), we re-write this relation
in the form
\begin{eqnarray}
  \Sigma^{\text{int}\,+}_{\sigma}(0)+ \Sigma^{\text{int}\,+}_{\bar{\sigma}}(0)=U\, .
\end{eqnarray}
Therefore, $\im \Sigma^{\text{int}}(0)=0$, in agreement with the
unitarity condition. Sending now the Zeeman field on the dot to
zero, yields the result $2\re\Sigma^{\text{int}}(0)=U$, which agrees
with the real part of Eq. (\ref{SYM0}). Thus the EOM result for
the real part of $\Sigma^{\text{int}}$ on the Fermi energy does not
have a discontinuity. We hence conclude that the EOM techniques
failure at the symmetric point is confined to the imaginary part
of the interacting self-energy alone and to the symmetric point
alone.

Our considerations in this subsection are confined to $\omega =0$,
and therefore do not allow us to investigate the Friedel sum-rule
easily. We carry out such an analysis for the infinite-$U$ case
below. Alternatively, one may attempt, as has been done in
Ref.~~\onlinecite{Bulka03}, to \emph{impose} the Friedel sum-rule
on the EOM result, assuming that Eq.~(\ref{FRIEDEL}) holds,  with
$\tilde{n}_{d\sigma}\equiv\qav{n_{d\sigma}}$.  This is a
dangerous procedure, which leads in some cases to un-physical
results, as is demonstrated in Sec.~\ref{sec:physUinf}.

\section{Exact solution in the $U\to\infty$ limit\label{sec:Uinf}}
In this section we present an exact solution of the
self-consistently truncated  EOM, and obtain the dot Green
function, in the limit $U\rightarrow\infty$. For this solution, we
assume that the bandwidth $D_\sigma$ is larger than the other
energies in the problem (except $U$). In the next section we use
this function to calculate several physical quantities, and
compare the results with those of the Bethe \emph{anstaz}
technique and other calculations. For simplicity, we also assume
that the non-interacting self-energy may be approximated by an
energy-independent resonance width, i.e.,
\begin{eqnarray}
\Sigma^{\pm}_{\sigma}=\mp i\Gamma_{\sigma}\ ,\label{ASSU}
\end{eqnarray}
for all the energies in the band, $-D_\sigma < \omega < D_\sigma$
(The extension to the case where there is also an
energy-independent real part to the self-energy is
straightforward). This assumption is certainly reasonable for a
range of energies near the center of the band. However, using it
for the whole band introduces corrections of order
$|\omega/D_\sigma|$, thus restricting the solution to
$|\omega/D_\sigma|\ll 1$. Then, Eq. (\ref{eq:GfiniteU}) for the
Green function, together with the definitions (\ref{u}),
(\ref{eq:PQselfcon}), and (\ref{eq:defThermal}), takes the form
\begin{eqnarray}
\bigl [\mathcal{G}_{\sigma}^{\pm}(\omega )\bigr
]^{-1}G_{\sigma}^{\pm}(\omega+\sigma h)=1-\qav{n_{d\bs}}
-P^{\pm}_{\bs}(\omega +\bs h) ,\label{GGI}
\end{eqnarray}
where
\begin{eqnarray}
\mathcal{G}_{\sigma}^{\pm}(\omega )&=&\bigl [ \omega-\epsilon_{0}\pm
i\Gamma_{\sigma}-I^{\pm}_{\bs}(\omega +\bs
h)\nonumber\\
&\mp&i\bigl (\Gamma_{\bs}+\Gamma_{\sigma}\bigr )P^{\pm}_{\bs
}(\omega +\bs h )\bigr ]^{-1} \, .\label{GI}
\end{eqnarray}
Note the shift of energies by $\sigma h$, compared to Eq.
(\ref{eq:GfiniteU}). The function $I_{\sigma}$ introduced here
contains the Kondo singularity, \cite{footaa}
\begin{eqnarray}
  I_{\sigma}^{\pm}(\omega)&=&\frac{\Gamma_{\sigma}}{\pi}\int_{-D_{\sigma}}^{D_{\sigma}}
  d\omega'   \frac{f(\omega') }{\omega\pm i\eta -\omega '}\, .\label{III}
\end{eqnarray}
To order $\mathcal{O}(\omega /D_{\sigma})$, one has
\begin{eqnarray}
\pi I_\sigma^{\pm} (\omega)/\Gamma_\sigma = -\Psi\left(\frac{1}{2}
\mp \frac{i \omega}{2 \pi T} \right ) + \ln  \frac{D_{\sigma}}{2
\pi T} \mp \frac{i\, \pi}{2} \, , \label{eq:Iin}
\end{eqnarray}
where $\Psi$ is the digamma function. Equation (\ref{GGI}) for the
Green function also contains the function $P_\sigma$, which, using
the assumption (\ref{ASSU}), is given by
\begin{eqnarray}
P^{\pm}_\sigma(\omega)  = \frac{\Gamma_{\sigma}}{\pi}
\int_{-D_{\sigma}}^{D_{\sigma}} d\omega ' \frac{f(\omega')\,
G^{\mp}_{\sigma}(\omega')}{\omega\pm i\eta-\omega'} \, d \omega' \, . \label{Papprox}
\end{eqnarray}
Physically, all the integrals which contain $\Sigma_\sigma$ must
be calculated between $-D_\sigma$ and $D_\sigma$, and the
resulting Green function is calculated only for energies inside
the band, $|\omega| <D_\sigma$. However, the integral
$P^{\pm}_\sigma(\omega)$ converges even when one takes the limit
$D_\sigma \to \infty$, because $G_\sigma(\omega) \sim 1/\omega$ at
large $|\omega|$ [see e.g. Eq. (\ref{GGI})]. If $D_\sigma$ is
sufficiently large, so that this asymptotic behavior becomes
accurate and since $f(\omega) \approx 1$ for $\omega < -D_\sigma$,
it is convenient to extend the range of this integral (and all the
related integrals below, unless otherwise specified) to the range
$-\infty<\omega <\infty$. This introduces errors of order
$\Gamma_\sigma/D_\sigma$ or $\omega/D_\sigma$ in the results,
which we neglect.

We have thus found that the equation for $G_{\sigma}^{+}$ involves
an integral containing the function $G^{-}_{\bs}$, and thus the
two functions $G^{+}_{\sigma}$ and $G^{-}_{\bs}$ are coupled. In
addition, the occupations $\qav{n_{d\sigma}}$ have to be
determined self-consistently from the Green functions themselves.
Our solution for the Green function follows the method introduced
in Refs.~~\onlinecite{Appelbaum69both} and
~\onlinecite{BloomfieldHamann67}. This method allows one to turn
the integral equations into algebraic ones, at the cost of
additional quantities which have to be determined
self-consistently from the Green function. First, one introduces
the functions
\begin{widetext}
\begin{eqnarray}
  \Phi_{\sigma}(z)&=&z-\epsilon_{0}+i\Gamma_{\sigma}-I_{\bs}(z+\bs
  h)-i\bigl (\Gamma_{\bs}+\Gamma_{\sigma}\bigr
  )\frac{\Gamma_{\bs}}{\pi}\int d\omega '\frac{f(\omega')
  G^{-}_{\bs}(\omega ')}{z+\bs h -\omega '}\ ,\nonumber\\
  \widetilde{\Phi}_{\sigma}(z)&=&z-\epsilon_{0}-i\Gamma_{\bs}-I_{\sigma}(z+\sigma
  h)+i\bigl (\Gamma_{\bs}+\Gamma_{\sigma}\bigr
  )\frac{\Gamma_{\sigma}}{\pi}\int d\omega '\frac{f(\omega')
  G^{+}_{\sigma}(\omega ')}{z+\sigma h -\omega '}\, .\label{PSI}
\end{eqnarray}
Note that $\Phi_{\sigma}^{+}(\omega )$ is identical to $\bigl
[\mathcal{G}_{\sigma}^{+}(\omega )\bigr ]^{-1}$ and
$\widetilde{\Phi}^{-}_{\sigma}(\omega )\equiv \bigl [\mathcal{
G}^{-}_{\bs}(\omega )\bigr ]^{-1}$, while
$\Phi_{\sigma}^{-}(\omega )$ and
$\widetilde{\Phi}^{+}_{\sigma}(\omega )$ are  \emph{different}
from $\bigl [\mathcal{G}_{\sigma}^{-}(\omega )\bigr ]^{-1}$ and
$\bigl [\mathcal{ G}^{+}_{\bs}(\omega )\bigr ]^{-1}$,
respectively. The knowledge of $\Phi^{\pm}_{\sigma}$ is sufficient
to determine $G^{-}_{\bs}$, since
\begin{eqnarray}
  \Phi_{\sigma}^{+}(\omega )-\Phi_{\sigma}^{-}(\omega
  )=2i\Gamma_{\bs}f(\omega +\bs h)\bigl [1+i\bigl
  (\Gamma_{\bs}+\Gamma_{\sigma}\bigr )G^{-}_{\bs }(\omega +\bs h
  )\bigr ]\, .\label{DIF1}
\end{eqnarray}
Similarly, the functions $\widetilde{\Phi}^{\pm}_{\sigma}$
determine $G^{+}_{\sigma}$, through the relation
\begin{eqnarray}
  \widetilde{\Phi}_{\sigma}^{+}(\omega
  )-\widetilde{\Phi}_{\sigma}^{-}(\omega )=2i\Gamma_{\sigma}f(\omega
  +\sigma h)\bigl [1-i\bigl (\Gamma_{\bs}+\Gamma_{\sigma}\bigr
  )G^{+}_{\sigma }(\omega +\sigma h )\bigr ]\, .\label{DIF2}
\end{eqnarray}
Returning now to Eq. (\ref{GGI}), we eliminate the Green functions
by using Eqs. (\ref{DIF1}) and (\ref{DIF2}), and the functions $P$
by using the definitions (\ref{PSI}). In this way we find
\begin{eqnarray}
  \Phi^{+}_{\sigma}(\omega
  )\frac{\widetilde{\Phi}^{+}_{\sigma}(\omega )-
  \widetilde{\Phi}^{-}_{\sigma}(\omega )}{2i\Gamma_{\sigma}f(\omega
  +\sigma h)}&=&{\rm X}_{\sigma}^{+}(\omega )\ , \ \ \
  \widetilde{\Phi}^{-}_{\sigma}(\omega
  )\frac{\Phi^{+}_{\sigma}(\omega )- \Phi^{-}_{\sigma}(\omega
  )}{2i\Gamma_{\bs}f(\omega +\bs h)}=\widetilde{\rm
  X}_{\sigma}^{-}(\omega )\,  ,\label{TTWW}
\end{eqnarray}
where
\begin{eqnarray}
  {\rm X}_{\sigma}(z)&=&-i(\Gamma_{\sigma}+\Gamma_{\bs})(1-\qav{
  n_{d\bs}})+z
  -\epsilon_{0}+i\Gamma_{\sigma}-I^{}_{\bs}(z+\bs h)\,
 ,\nonumber\\
  \widetilde{\rm
  X}_{\sigma}(z)&=&i(\Gamma_{\sigma}+\Gamma_{\bs})(1-\qav{
  n_{d\sigma}})+z -\epsilon_{0}-i\Gamma_{\bs}-I^{}_{\sigma}(z
  +\sigma h)\, .\label{X}
\end{eqnarray}
So far, we have not achieved much simplification over the original
problem at hand. However, noting that
\begin{eqnarray}
{\rm X}^{+}_{\sigma}(\omega )-{\rm X}^{-}_{\sigma}(\omega
)&=&2i\Gamma_{\bs}f(\omega +\bs h)\ , \ \ \
\widetilde{X}^{+}_{\sigma}(\omega
)-\widetilde{X}^{-}_{\sigma}(\omega )=2i\Gamma_{\sigma}f(\omega
+\sigma h)\, ,
\end{eqnarray}
Eqs. (\ref{TTWW}) yield the remarkable result
\begin{eqnarray}
\Phi_{\sigma}^{+}(\omega )\, \widetilde{\Phi}_{\sigma}^{+}(\omega
)-{\rm X}^{+}_{\sigma}(\omega )\, \widetilde{\rm
X}^{+}_{\sigma}(\omega )=\Phi_{\sigma}^{-}(\omega
)\, \widetilde{\Phi}_{\sigma}^{-}(\omega )-{\rm
X}^{-}_{\sigma}(\omega )\, \widetilde{\rm X}^{-}_{\sigma}(\omega )\,
.\label{BREAK}
\end{eqnarray}
Therefore, the combination
\begin{eqnarray}
R(z)\equiv\Phi_{\sigma} (z)\, \widetilde{\Phi}_{\sigma}(z)-{\rm
X}_{\sigma}(z)\, \widetilde{\rm X}_{\sigma}(z)
\end{eqnarray}
is \emph{non-singular} across the real axis. In fact, the only
singular point of this combination is at $z=\infty$. This means
that $R(z)$ can be written as a polynomial with non-negative
powers of $z$. Moreover, since $\Phi$, $\widetilde{\Phi}$,  X, and
$\widetilde{\rm X}$ grow only linearly as $z\rightarrow\infty$,
that polynomial includes only two terms, $r_{0}+r_{1}z$. The
details of this calculation are given in Appendix \ref{ASEX}. The
result (\ref{BREAK}) allows one to express the (unknown) functions
$\Phi$ and $\widetilde{\Phi}$ in terms of the (known) functions
${\rm X}$, $\widetilde{\rm X}$ and $R$,
\begin{eqnarray}
\frac{\Phi_{\sigma}^{+}}{\Phi_{\sigma}^{-}}=\frac{R+{\rm
X}^{+}_{\sigma}\widetilde{\rm X}^{-}_{\sigma}}{R+{\rm
X}^{-}_{\sigma}\widetilde{\rm X}^{-}_{\sigma}}\equiv \rm
H_{\sigma}(\omega )\ ,\ \
\frac{\widetilde{\Phi}_{\sigma}^{+}}{\widetilde{\Phi}_{\sigma}^{-}}=\frac{R+{\rm
X}^{+}_{\sigma}\widetilde{\rm X}^{+}_{\sigma}}{R+{\rm
X}^{+}_{\sigma}\widetilde{\rm X}^{-}_{\sigma}}\equiv
\widetilde{\rm H}_{\sigma}(\omega )\ .\label{HH}
\end{eqnarray}
This reduces our problem into two independent \emph{linear
Reimann-Hilbert problems},  for which a rigorous solution is
available\cite{BloomfieldHamann67,Muskhlishvili53},
\begin{eqnarray}
\Phi_{\sigma} (z)=(z-a)e^{M_{\sigma}(z)}\, , && \ \
M_{\sigma}(z)=\int \Bigl (-\frac{d\omega}{2\pi i}\Bigr ) \frac{\ln
{\rm H}_{\sigma}(\omega )}{
z-\omega }\, ,\nonumber\\
\widetilde{\Phi}_{\sigma}(z)=(z-\widetilde{a})e^{\widetilde{M}_{\sigma}(z)}\,
, && \ \ \widetilde{M}_{\sigma}(z)=\int \Bigl
(-\frac{d\omega}{2\pi i}\Bigr ) \frac{\ln \widetilde{\rm
H}_{\sigma}(\omega )}{ z-\omega }\, .\label{SOL}
\end{eqnarray}
\end{widetext}
This is a valid solution as long as $\ln {\rm H}(\omega)$ and $\ln
\widetilde{\rm H}(\omega)$ can be chosen to be continuous in
$\omega$ and to vanish at both ends of the integration
interval.\cite{foot3} All the cases studied in this paper obey
this requirement, although we could not prove the absence of
solutions other than (\ref{SOL}) for a general case with no spin
symmetry. The form of the polynomial prefactors, $(z-a)$ and $(z-
{\tilde a})$, in Eq.~\eqref{SOL} is dictated by the fact that the
leading term in $\Phi_{\sigma}(z)$ and
$\widetilde{\Phi}_{\sigma}(z)$ must be $z$ [see Eqs.~\eqref{PSI}].
The determination of the coefficients $a$ and $\widetilde{a}$, as
well as other self-consistent quantities, is detailed in Appendix
\ref{RH}.

\section{Physical properties in the $U\to\infty$ limit\label{sec:physUinf}}
\begin{figure}
\includegraphics[width=8cm]{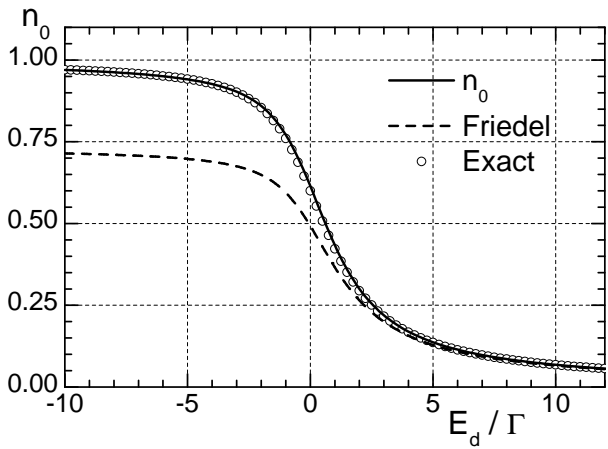}
  \caption{The zero-temperature occupation number $n_0$ as function
  of the renormalized energy  $E_d$ calculated self-consistently by
  the EOM method  (solid line), and from the Friedel sum rule, Eq.
  (\ref{FRIEDEL}) (dashed line). Open circles show the exact Bethe
  \emph{ansatz} results\cite{WiegmannC83Ogievetski83}.
   \label{fig:n0}}
\end{figure}

Once the functions $\Phi$ and $\widetilde{\Phi}$ are found, then
the Green function is determined from Eq. (\ref{DIF1}) or Eq.
(\ref{DIF2}). This knowledge enables us to compute various
physical quantities, and compare them with the results of other
calculations. The first quantity we consider is the total
occupation on the dot,
$n_0\equiv\qav{n_{d\spinup}+n_{d\spindown}}$, at zero temperature.
This calculation is carried out for the spin-symmetric case, $h=0$
and spin-independent self-energy. We also denote $D \equiv
D_\sigma,~\Gamma \equiv \Gamma_\sigma$. The result is plotted as
function of $ E_d/\Gamma$, where
\begin{eqnarray}
  E_{d} \equiv \epsilon_0 +(\Gamma/\pi) \ln (D/\Gamma)\, ,\label{Ed}
\end{eqnarray}
and is portrayed in Fig.~\ref{fig:n0} (full line). It agrees
within $3 \%$ with the exact universal curve $n_0(E_d/\Gamma)$, as
found from the Bethe \emph{ansatz}, \cite{WiegmannC83Ogievetski83}
(open circles). Thus, the EOM solution conforms with  Haldane's
scaling. \cite{Haldane78} On the other hand, the EOM solution
fails to satisfy the Friedel sum-rule, as has been already
discussed above. The total occupation calculated from Eq.
(\ref{FRIEDEL}) (dashed line) deviates systematically from the
self-consistent values, in particular in the Kondo regime ($n_0
\to 1$). Note that Eq. (\ref{ASSU}) implies that
$\tilde{n}_{d\sigma}= \qav{n_{d\sigma}}$ [see Eq. (\ref{ntild})].

As mentioned above, the Fermi-liquid relations are connected with
unitarity. In particular, at zero temperature the linear
conductance of a symmetrically-coupled quantum dot is given by
\cite{NgLee88GlazmanRaikh88,Meir91ANDGF} $-2(e^2/h)\Gamma_\sigma
\im  G_\sigma(\omega=0)$. Thus, the unitary limit $2e^2/h$ is
reached only if $\re [G^+_\sigma(0)]\to 0$ and the Fermi liquid
relation (\ref{eq:unitarity}) holds. As implied by the dashed line
in the figure, the first of these criteria is not obeyed by the
self-consistent solution.

Next we consider the local spin susceptibility on the dot. When
the leads are non-magnetic, this quantity is given by $\chi=
\frac{1}{2}(g \mu_B)^2 \partial
\qav{n_{d\spindown}-n_{d\spinup}}/\partial h $, with $ h =-g \mu_B
B$. Here $B$ is the external magnetic field and $g \mu_B$ is the
gyromagnetic ratio of electrons in the quantum dot. (The spin
susceptibility of the leads adds to the local susceptibility the
usual Pauli term, and $\mathcal{O}(\Gamma_{\sigma}/D_{\sigma})$
corrections. \cite{Hewson93}{}) We have calculated
$\chi(E_d/\Gamma,T)$ by differentiating the self-consistent
equations for $\qav{n_{d\sigma}}$  with respect to $ h $, and
evaluating the integrals numerically. This procedure is similar to
the one which has been used for the Wolff model in
Ref.~~\onlinecite{Appelbaum69both}b, but is free from numerical
accuracy problems reported there.

\begin{figure}
 \includegraphics[width=8cm]{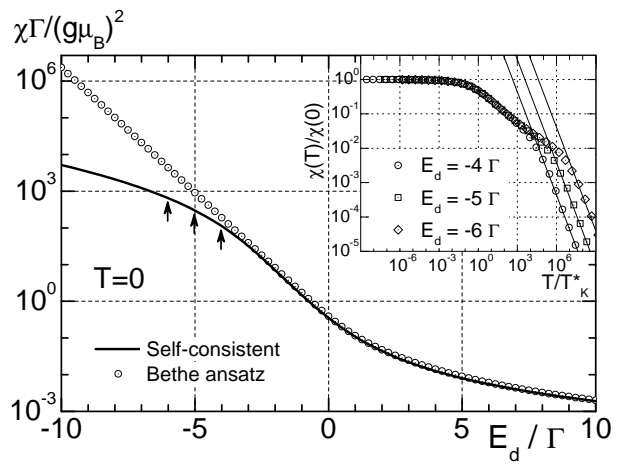}
 \caption{\label{fig:chi}The local spin susceptibility $\chi$ as  function
 of the renormalized energy level $E_d$, for $T=0$. The EOM result
 (solid line) is close to  the Bethe \emph{ansatz}
 one \cite{WiegmannC83Ogievetski83} (circles), except for large
 negative values of $E_d/\Gamma$. Inset: The scaling of the susceptibility with
 $T/T_{K}^{\ast}$ at three fixed energy level positions
 (marked by arrows in the main graph).
 The solid lines indicate the high-temperature asymptotic behavior,
 $\chi(T)=(g \mu_B)^{2}/(6 T)$.}
\end{figure}

The zero-temperature susceptibility derived from the EOM is found
to be in a good \emph{quantitative} agreement with the Bethe
\emph{ansatz} results in the mixed valence ($|E_d/\Gamma| \lesssim
1$) and empty orbital ($E_d \gg \Gamma$) regimes, as is shown in
Fig.~\ref{fig:chi}. In the  local moment regime, $E_d \ll -
\Gamma$, a screening cloud is expected to form due to the Kondo
effect at\cite{Hewson93,Haldane78} $T< T_K \sim \Gamma e^{\pi
E_d/(2\Gamma)}$, leading to a crossover from a high-temperature
Curie law, $\chi \sim (g \mu_B)^2/T$, to a finite ground state
value, $\chi \sim (g \mu_B)^2/T_K$. The latter is underestimated
by our self-consistent solution, as is manifested by the deviation
from the exact Bethe \emph{ansatz} results depicted in
Fig.~\ref{fig:chi}.

Indeed, had we defined the Kondo temperature through the inverse
of the zero-temperature susceptibility, we would have found that
the EOM method \emph{overestimates} that temperature. However,
within EOM, the relevant energy scale is determined from the
leading (real) terms in the denominator of the Green function,
i.e., by the temperature at which  the real part of $\mathcal{G}$,
Eq. (\ref{GI}), vanishes. Using for $I(z)$, Eq. (\ref{eq:Iin}),
the approximate form \cite{Theumann69} $I_{}(z) \approx
-(\Gamma/\pi) \log \left [(z + i \kappa T  )/D \right]$ where
$\kappa$ is a number of order unity, we find that the leading
terms are
\begin{eqnarray}
  \omega -E_{d}+\frac{\Gamma}{\pi}\ln\frac{D}{\Gamma}+
  \frac{\Gamma}{\pi}\ln\frac{\sqrt{\omega^{2}+
  \kappa^{2} T^{2}}}{D}\, ,
\end{eqnarray}
yielding for the temperature scale
\begin{eqnarray}
  T_K^{\ast}\sim \Gamma e^{\pi E_d/\Gamma}\, ,\label{TSTAR}
\end{eqnarray}
such that the leading terms are
\begin{eqnarray}
  \omega +\frac{\Gamma}{\pi} \ln \sqrt{(\omega/T_K^{\ast})^2+(\kappa
  T/T_K^{\ast})^2}\, .\label{DOM}
\end{eqnarray}
The logarithm in Eq. (\ref{DOM}) dominates the properties of the
solution close to the Fermi energy at temperatures $T \lesssim
T_K^{\ast}$. The same energy scale $T_K^{\ast}$ has been
determined from the analysis of the truncated EOM in
Refs.~~\onlinecite{Lacroix81} and
~\onlinecite{EntinAharonyMeir04}. Note that $T_K^{\ast}$ is
\emph{smaller} than the true Kondo temperature $T_K$.

The local spin susceptibility at finite temperatures, calculated
from our EOM solution, is shown in the inset of
Fig.~\ref{fig:chi}. Indeed, $\chi(T)/\chi(0)$ scales with
$T/T_K^{\ast}$, but instead of crossing over to the Curie law, a
region of intermediate behavior in which $\chi \sim T^{x}$ with
$-1< x<0$ is observed. The high-temperature asymptotic $\chi(T)
\sim 1/T$ is approached only for $T\gtrsim\Gamma$. Such a behavior
in the intermediate temperature range $T_K^{\ast} < T < \Gamma$ is
not supported by NRG or Bethe \emph{ansatz}
calculations\cite{Hewson93}, which scale with $T_K$,~\cite{wolfle}
and has thus to be attributed to the deficiency of the EOM method.

\begin{figure}
  \includegraphics[width=8cm]{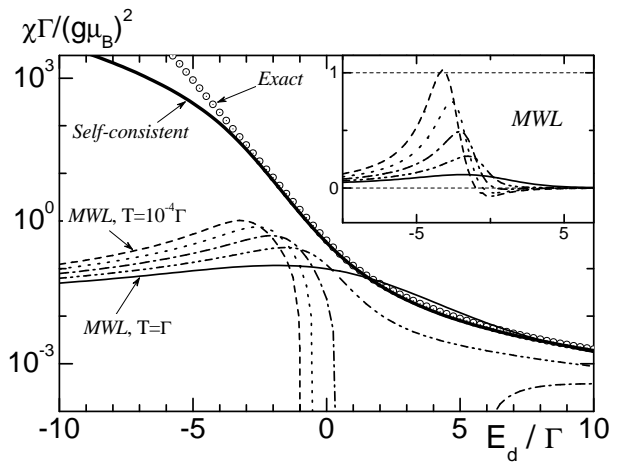}
  \caption{\label{fig:Meir}Data from Fig.~\ref{fig:chi} compared to
  the spin susceptibility given by the MWL approximation in the
  temperature range from $T=1 \Gamma$ down to $T=10^{-4} \Gamma$.
  The inset shows the MWL susceptibilities only, on a linear scale,
  allowing for the negative values.}
\end{figure}

In contrast, neither the Lacroix approximation, as implemented in
Ref.~~\onlinecite{EntinAharonyMeir04}, nor the
MWL\cite{Meir91ANDGF} approximate Green function leads to
comparable results when used to calculate the local spin
susceptibility. The Lacroix approximation becomes intrinsically
inconsistent at finite magnetic fields, since it results in a
logarithmic divergence of $G_{\sigma}(z)$ as $z\to \sigma h$. Even
when ignoring this inconsistency, the zero-temperature spin
susceptibility calculated in that approximation attains  negative
and divergent values regardless of the quantum dot parameters. The
MWL approximation leads to finite, but quite unphysical values of
$\chi$ for $T < \Gamma$, as we demonstrate in Fig.~\ref{fig:Meir}.
That approximation gives reasonable results at  $T=\Gamma$:  the
susceptibility follows roughly the Curie law $\approx
1/T(=1/\Gamma$)  or the zero-temperature value, whichever is
smaller. At lower temperatures one would have expected a gradual
increase in the susceptibility in the Kondo region. Instead, a
window of a \emph{negative susceptibility} opens, which is widened
as the temperature is decreased. At strictly zero temperature
$\chi$ is negative for all values of $E_d$. This example shows the
necessity of using  the full self-consistency of the EOM solution
in order to obtain the qualitatively correct behavior.

\begin{figure}
  \includegraphics[width=8cm]{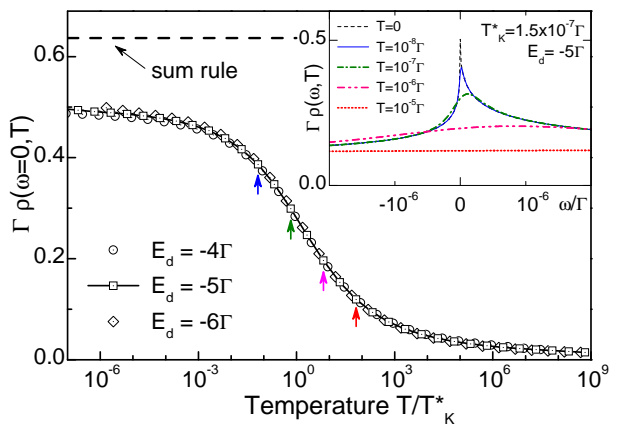}
  \caption{\label{fig:DOS}(Color online) The scaling of the local density of
  states at the Fermi level as function of the reduced temperature
  $T/T^{\ast}_K$, is demonstrated by plotting
   $\rho(\omega =0,T)$ for three different values of the
   renormalized single-electron energy on the dot, $E_d$ [Eq. (\ref{Ed})]. The dashed
  line at $\rho=2/(\pi \Gamma)$ corresponds to the limit dictated by
  the Friedel sum rule at $E_{d}/\Gamma \ll -1$. The inset shows the
  melting of the Kondo peak as the temperature is increased at fixed
  $E_d/\Gamma=-5$. Temperature values represented in the inset are
  marked by arrows in the main graph. }
\end{figure}

Finally, we examine the local density of states on the dot, given
by
\begin{eqnarray}
\rho(\omega,T)\equiv -\im \sum_\sigma G^{+}_{\sigma}(\omega)/\pi
\, ,
\end{eqnarray}
as calculated from the EOM. The inset of Fig. ~\ref{fig:DOS} shows
the Kondo peak at temperatures $T \lesssim T_{K}^{\ast}$. The
local density of states at the Fermi energy follows the universal
temperature dependence, as can be seen in Fig. ~\ref{fig:DOS},
with the same scaling factor $T^{\ast}_K$ as the spin
susceptibility (compare to the inset of Fig.~\ref{fig:chi}). The
appearance  of a single scale determining the low energy
properties of the system is another hallmark\cite{Hewson93} of the
Kondo effect which is captured by the fully self-consistent EOM
technique.

\section{Perturbation expansion in the dot-network couplings}
\label{PT}
The EOM technique is unfortunately not a systematic
expansion. It is therefore very interesting to compare its results
with those given by a direct expansion. Here we expand the EOM
Green function up to second-order in the dot-network couplings
$J_{n{\sigma}}$. For compactness, we confine ourselves to the case
$U \to \infty$, and assume for simplicity that the non-interacting
self-energy has just an energy-independent imaginary part, i.e.,
$\Sigma_{\sigma}(\omega)=-i \Gamma_{\sigma}$. By comparing with
the direct perturbation theory expansion \cite{Sivan96,Gefen04} we
find that, up to second-order in $J_{n{\sigma}}$, our EOM result
for the dot Green function is exact. On the other hand, the Green
function derived in Ref.~~\onlinecite{Meir91ANDGF} (which does not
include all correlations resulting from the truncated EOM)
violates the second-order perturbation theory result, and in fact,
predicts a Kondo anomaly in this order (which should not be
there).

\begin{figure}
  \includegraphics[width=8cm]{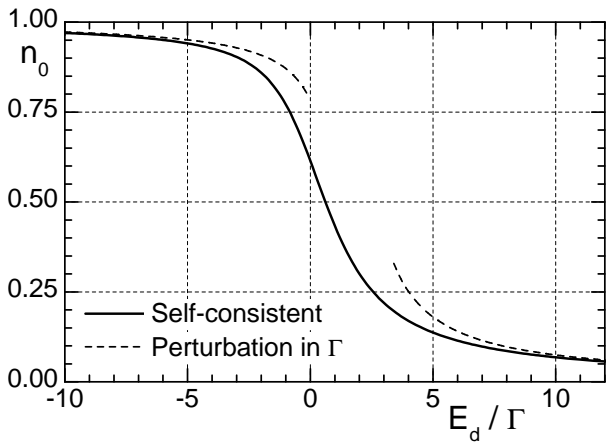}
  \caption{The equilibrium occupation number
  $n_0=\sum_{\sigma}\qav{n_{\sigma}}$ for the same
  parameters as in Fig.~\ref{fig:n0}, and $D=100\Gamma$ (since
  second-order perturbation theory \emph{does not} scale with
  $E_{d}$, the bandwidth has to be specified), calculated from  the
  self-consistently truncated EOM (solid lines), and by perturbation
  theory to first-order in $\Gamma$, Eqs. (\ref{DELn0}) and
  \eqref{Gefen} (dashed line). The perturbational result diverges at
  $\epsilon_0=0$.
  \label{fig:nPT}}
\end{figure}

The $U \to \infty$ Green function of the EOM is given by Eqs.
(\ref{GGI}), (\ref{GI}), (\ref{III}), and (\ref{Papprox}). Since
the non-interacting self-energy $\Sigma_{\sigma}$, and
consequently the width $\Gamma_{\sigma}$ are second-order in the
coupling $J_{n\sigma}$, the expansion of the Green function reads
\begin{eqnarray}
  G^{(0)}_{\sigma}(z)   =  \frac{\delta n_{\bar{\sigma}}^{(0)}}{z-
  \epsilon_\sigma} \, ,  \label{G0}
\end{eqnarray}
and
\begin{eqnarray}
  G_{\sigma}^{(2)}(z) & = & \frac{\delta
  n_{\bar{\sigma}}^{(2)}-P^{(2)}_{\bar{\sigma}}(z_1)}
  {z-\epsilon_{\sigma}}\nonumber\\
  &+&G_{\sigma}^{(0)}(z) \frac{I_{\bar{\sigma}}(z_1) +
  \Sigma_{\sigma}(z)}{z-\epsilon_{\sigma}}  \, , \label{GSEC}
\end{eqnarray}
where we have denoted
\begin{eqnarray}
  \qav{n_{\bar{\sigma}}}^{}\equiv 1- \delta n_{\bar{\sigma}}^{}\ ,\
  \ \epsilon_{\sigma}=\epsilon_{0}+\sigma h  \label{DELn}
\end{eqnarray}
and where the superscript $(k)$ denotes the contribution of order
$k$ in the couplings. The function $P^{(2)}$ is found by using the
zeroth-order of the Green function in Eq. (\ref{Papprox}),
\begin{eqnarray}
  P^{(2)}_{\sigma}(z)&= & G_{\sigma}^{(0)}(z) \bigl [ I_{\sigma}(z)-
  I_{\sigma}^{-}(\epsilon_{\sigma})\nonumber\\
  &-&f(\epsilon_{\sigma}) \Sigma_{\sigma}(z) +f(\epsilon_{\sigma})
  \Sigma^{-}_{\sigma}(\epsilon_{\sigma})\bigr ] \ .
 \end{eqnarray}
Using the identities
\begin{eqnarray}
  I_{\sigma}^{-}(\epsilon_{\sigma})-f(\epsilon_{\sigma})
  \Sigma^{-}_{\sigma}(\epsilon_{\sigma}) = \re \bigl [
  I_{\sigma}(\epsilon_{\sigma})-
   f(\epsilon_{\sigma}) \Sigma_{\sigma}(\epsilon_{\sigma}) \bigr
   ]
\end{eqnarray}
and
\begin{eqnarray}
  G_{\bar{\sigma}}^{(0)}(z_1)=\delta n_{\sigma}^{(0)}
  G_{\sigma}^{(0)}(z) / \delta n_{\bar{\sigma}}^{(0)}\, ,
\end{eqnarray}
Eq. (\ref{GSEC}) takes the form
\begin{widetext}
\begin{align}
  G_{\sigma}^{(2)}(z) =\frac{\delta
n_{\bar{\sigma}}^{(2)}}{z-\epsilon_{\sigma}}+\frac{ -\delta
n_{\sigma}^{(0)} \left[
I_{\bar{\sigma}}(z_1)-f(\epsilon_{\bar{\sigma}})
\Sigma_{\bar{\sigma}}(z_1) -
  \re \{ I_{\bar{\sigma}}(\epsilon_{\bar{\sigma}})-
  f(\epsilon_{\bar{\sigma}}) \Sigma_{\bar{\sigma}}(\epsilon_{\bar{\sigma}})
  \}\right]
 +\delta n_{\bar{\sigma}}^{(0)} \left [
 I_{\bar{\sigma}}(z_1)+ \Sigma_{\sigma}(z)\right ]
}{(z-\epsilon_\sigma)^2} \label{eq:G2ready} \, .
\end{align}
\end{widetext}
In order to complete the second-order calculation, we need to find
the occupation  numbers $\delta n_{\sigma}^{(0)}$ and $\delta
n_{\sigma}^{(2)}$ [see Eq. (\ref{DELn})]. From Eq. (\ref{G0}) we
find
\begin{align}
  \qav{n_{\sigma}}^{(0)} & = \frac{f(\epsilon_\sigma) [
  1-f(\epsilon_{\bar{\sigma}}) ]}{1-f(\epsilon_\sigma)
  f(\epsilon_{\bar{\sigma}})}\, . \label{DELn0}
\end{align}
The second-order correction to the occupation number, $\delta
n_{\sigma}^{(2)}\equiv -\qav{n_{\sigma}}^{(2)}$, is obtained by
integrating over the Fermi function multiplied by the second-order
correction to the density-of-states, $\rho^{(2)}_{\sigma}(\omega
)$. The latter reads
\begin{widetext}
\begin{align}
\rho^{(2)}_{\sigma}(\omega ) \equiv
  \frac{G_{\sigma}^{+(2)}(\omega)-G_{\sigma}^{-(2)}(\omega)}{-2 \pi i}
= & \delta (\omega -\epsilon_{\sigma})\Bigl (\delta
n^{(2)}_{\bar{\sigma}}+ \left [
\qav{n_{\sigma}}^{(0)}-\qav{n_{\bar{\sigma}}}^{(0)} \right]
\frac{\partial\re
I_{\bar{\sigma}}(\epsilon_{\bar{\sigma}})}{\partial\epsilon_{0}}\Bigr
) -  \delta '(\omega -\epsilon_{\sigma})\delta
n^{(0)}_{\bar{\sigma}}\re
I_{\bar{\sigma}}(\epsilon_{\bar{\sigma}})
\nonumber\\
+& \frac{
 \Gamma_{\sigma} +
    \left( \Gamma_{\bar{\sigma}}
-\Gamma_{\sigma}\right ) \qav{n_{\bar{\sigma}}}^{(0)}
 +
 \Gamma_{\bar{\sigma}} \, f(\omega-\epsilon_{\sigma}+\epsilon_{\bar{\sigma}})
\left[ \qav{n_{\sigma}}^{(0)} -  \qav{n_{\bar{\sigma}}}^{(0)}
  \right]
}{\pi (\omega -\epsilon_{\sigma})^{2}} \, .\label{RHO2}
\end{align}
Apart from the terms representing the second-order modifications
of the singularity at $\epsilon_{\sigma}$ [the first and second
members of Eq. (\ref{RHO2})], our result reproduces the one of
Ref.~~\onlinecite{Sivan96}, for the case where the width
$\Gamma_{\sigma}$ is spin-independent, i.e.,
$\Gamma_{\sigma}=\Gamma_{\bar{\sigma}}$. Note that  this density
of states  remains finite at all temperatures. The Kondo
divergence  of $\rho^{(k)}_{\sigma}(0 )$ in the limit  $T \to 0$
appears only at $k \ge 4$ [see Ref.~~\onlinecite{Sivan96}]. Using
the second-order correction to the density of states, we find
\begin{align}
\qav{n_{\sigma}}^{(2)} = \int f(\omega)\rho^{(2)}_{\sigma}(\omega
) d\omega =\left ( \qav{ n_{\bar{\sigma}}}^{(0)} -1 \right ) \re
\frac{\partial I_{\sigma}(\epsilon_{\sigma})}{\partial
\epsilon_{\sigma}} + \frac{\partial\qav{
n_{\sigma}}^{(0)}}{\partial\epsilon_{\sigma}} \re
I_{\bar{\sigma}}(\epsilon_{\bar{\sigma}})+\frac{\partial\qav{
n_{\sigma}}^{(0)}}{\partial\epsilon_{\bar{\sigma}}}  \re
I_{\sigma}(\epsilon_{\sigma})\ . \label{Gefen}
\end{align}
\end{widetext}
This result is identical to the $U\to \infty$ limit of Eq.~(5) in
Ref.~~\onlinecite{Gefen04} which was obtained by a direct
perturbation expansion. Figure \ref{fig:nPT} depicts the total
occupation, $n_0=\sum_{\sigma}\qav{n_{\sigma}}$, as function of
$\epsilon_{0}$, as found from the EOM technique, and as computed
from Eqs. (\ref{DELn0}) and \eqref{Gefen} to first order in the
width. The two curves differ by a few percents. The comparison
with the exact result is carried out in Sec.~\ref{sec:Uinf}.

\section{Conclusions}
\label{sec:conclusions} We have presented a solution for the Green
function of an interacting quantum dot embedded in a general
non-interacting network. Our solution is derived within the EOM
technique, taking into account exactly \emph{all} resulting
correlations once those equations are truncated. We have tested
our solution by analyzing several limiting cases, and by comparing
several physical properties derived from that solution with other
results available by the Bethe \emph{ansatz} method and by NRG
computations.

We have found that the EOM Green function is
temperature-independent at the particle-hole symmetric point
(where $h=0$ and $2\epsilon_{0}+U=0$). We have found that this
deficiency of the EOM is related to a discontinuity in the
imaginary part of the interacting self-energy at that particular
point. However, this imaginary part obeys the Fermi liquid
unitarity requirement away from this special point, at zero
temperature. In contrast, even away from the particle-hole
symmetric point, the EOM result  fails to satisfy the Friedel
sum-rule deep inside the Kondo regime, as we have shown explicitly
in the infinite interaction limit.

Albeit these problems, the EOM solution reproduces faithfully the
low-temperature scaling of the spin susceptibility and the density
of states at the Fermi level, though with an energy scale
$T^{\ast}_{K}$ which differs from the true Kondo temperature
[\emph{ cf.} Eq. (\ref{TSTAR})]. Zero-temperature results are in
excellent agreement with the exact Bethe \emph{ansatz} solution,
except for the Kondo correlated regime. As the temperature is
raised, the EOM results become more and more quantitatively
correct, and approach high-temperature asymptotics known
rigorously from perturbation theory and NRG studies. We have
expanded the EOM Green function to second order in the dot-network
coupling and found an exact agreement with direct calculations by
perturbation theory. Most importantly, we have found that it is
crucial to include in the EOM solution \emph{all} the correlations
emerging from the truncated scheme. Ignoring part of these
correlations, as is ubiquitously done in such studies, results in
erroneous behaviors of various physical quantities.

Hence we conclude that the method examined in this paper can
provide a reasonable description of a quantum dot system over a
wide parameter range, provided that the self-consistency
conditions inherent to this technique are fully taken into
account.

%\begin{acknowledgments}
We thank A. Schiller for helpful comments. This project was
carried out in a center of excellence supported by the Israel
Science Foundation.
%\end{acknowledgments}

\appendix

\begin{widetext}
\section{EOM for the finite $U$ case\label{sec:appendixA}}
Here we extend the derivations of Refs.~~\onlinecite{Theumann69},
~\onlinecite{MamadaTakano70}, and
~\onlinecite{EntinAharonyMeir04}, carried out for an infinite
repulsive interaction,  to the case in which $U$ is finite. In
addition, we allow for  a Zeeman field, assuming that the
quantization axes  of the spins are the same on the dot and on the
leads (but the $g$-factors may be different). The Hamiltonian of
the our system is given in Eq.~\eqref{HAM}.

The Fourier transform of the EOM for the Green function defined in
Eq. (\ref{GREEN}) can be written in two alternative forms
\cite{Zubarev60} (both will be used in the following)
\begin{eqnarray}
z \qavens{A; B}_z  = \qav{\left [ A , B \right ]_{+}} +
\qavens{\left [ A , \ham \right ]_{-} ; B }_z = \qav{\left [ A , B
\right ]_{+}} - \qavens{A ; \left [ B , \ham \right ]_{-} }_z\
.\label{GENERALEOM}
\end{eqnarray}
It follows that the EOM for the dot Green function is
\begin{eqnarray}
\left[ z - \epsilon_0-\sigma h \right] G_\sigma  = && 1+ U
\qavens{ n^{}_{d\bar{\sigma}} d^{}_\sigma ; d_{\sigma}^{\dagger} }
- \sum_{n} J_{n \sigma} \qavens{ a^{}_{n \sigma} ;
d^\dagger_\sigma} \label{GdExact} \  ,
\end{eqnarray}
where $\bar{\sigma}$ is the spin direction opposite to $\sigma$.
Here and in the following we frequently omit for brevity explicit
indications of the argument $z$. We first consider the last term
on the right-hand-side of Eq. (\ref{GdExact}). The EOM for the
Green function appearing there is
\begin{eqnarray}
\left( z- \varepsilon_{n\sigma} \right)   \qavens{ a^{}_{n \sigma}
; d^\dagger_\sigma}  =    -\sum_m  J^{}_{nm} \qavens{ a^{}_{m
\sigma} ; d^\dagger_\sigma}-
 J^{\ast}_{n\sigma} G_{\sigma}  \
. \label{eq:cdExact}
\end{eqnarray}
Introducing the inverse matrix
\begin{eqnarray}
\mathcal{M}_{nm\sigma}(z) \equiv \left
[(z-\ham_{\text{net}}^{\sigma})^{-1}\right ]_{nm} \, ,
\label{MMATRIX}
\end{eqnarray}
where $\ham_{\text{net}}^{\sigma}$ is the part of $\ham_{\text{
net}}$ pertaining to the spin direction $\sigma$, we find
\begin{eqnarray}
\qavens{a^{}_{n\sigma}; d^{\dagger}_{\sigma}}
=-\sum_{m}\mathcal{M}^{}_{nm\sigma}(z)J^{\ast}_{m\sigma} \,
G_\sigma\ .\label{ad}
\end{eqnarray}
Note that $\mathcal{M}$ is the Green function matrix of the
network \emph{in the absence of the coupling to the dot}. Using Eq.
(\ref{ad}), we find that the last term on the right-hand-side of
Eq. (\ref{GdExact}) can be put in the form
\begin{eqnarray}
 -\sum_{n} J_{n \sigma} \qavens{ a^{}_{n \sigma} ;
d^\dagger_\sigma}=\Sigma_{\sigma}G_{\sigma}\ ,
\end{eqnarray}
where
\begin{eqnarray}
\Sigma_{\sigma}(z) \equiv \sum_{nm} J^{}_{n\sigma}
\mathcal{M}^{}_{nm\sigma} (z) J_{m\sigma}^{\ast}\label{Sigmadef}
\end{eqnarray}
is the self-energy of the dot Green function coming from the
coupling to the (non-interacting) network. Namely, it is the dot
self-energy for the $U=0$ case\cite{EntinAharonyMeir04}. As such,
it can always be calculated, at least in principle (see for
example Refs.~~\onlinecite{EntinAharonyMeir04} and ~\onlinecite{Pascal05}).

We now turn to the interacting part of the EOM for the dot Green
function [the second term on the right-hand-side of Eq.
(\ref{GdExact})]. The EOM for the 4-operator Green function
appearing there reads
\begin{eqnarray}
\left[ z - \epsilon_0 - \sigma h - U  \right] \qavens{
n^{}_{d\bar{\sigma}} d^{}_\sigma ; d_{\sigma}^{\dagger} }  &=&
\qav{n^{}_{d\bar{\sigma}}} -\sum_{n} \Bigl [J_{n\sigma}
\qavens{a^{}_{n\sigma} n^{}_{d\bar{\sigma}};
d^\dagger_\sigma}\nonumber\\
&+&J_{n\bar{\sigma}} \qavens{d^{\dagger}_{\bar{\sigma}}
a^{}_{n\bar{\sigma}} d^{}_{\sigma}; d^\dagger_\sigma} -
J^{\ast}_{n\bar{\sigma}} \qavens{a_{n\bar{\sigma}}^{\dagger}
d^{}_{\bar{\sigma}} d^{}_{\sigma}; d^\dagger_\sigma} \Bigr ]\, ,
\label{GammadExact}
\end{eqnarray}
and gives rise to three new 4-operators Green functions (on the
right-hand-side here). Their EOM are
\begin{eqnarray}
\left[ z- \varepsilon_{n \sigma} \right] \qavens{a^{}_{n\sigma}
n^{}_{d\bar{\sigma}}; d^\dagger_\sigma}  = &-&\sum_m J_{nm}
\qavens{a^{}_{m\sigma}  n^{}_{d\bar{\sigma}}; d^\dagger_\sigma} -
J^{\ast}_{n\sigma} \qavens{ n^{}_{d\bar{\sigma}} d^{}_\sigma ;
d_{\sigma}^{\dagger} } \nonumber \\ & - & \sum_{m} \left [
J_{m\bar{\sigma}} \qavens{a^{}_{n\sigma}
d^{\dagger}_{\bar{\sigma}} a^{}_{m \bar{\sigma}} ;
d^{\dagger}_\sigma} - J^{\ast}_{m \bar{\sigma}} \qavens{a_{m
\bar{\sigma}}^\dagger d^{}_{\bar{\sigma}} a^{}_{n\sigma} ;
d^{\dagger}_\sigma} \right ] \, , \label{eq:Gamma1Exact}
\end{eqnarray}
\begin{eqnarray}
\left[ z_1-\varepsilon_{n\bar{\sigma}} \right]
\qavens{d^{\dagger}_{\bar{\sigma}} a^{}_{n\bar{\sigma}}
d^{}_{\sigma}; d^\dagger_\sigma}  = &&
\qav{d^\dagger_{\bar{\sigma}} a^{}_{n\bar{\sigma}}}  -\sum_m
J_{nm} \qavens{d^{\dagger}_{\bar{\sigma}} a^{}_{m\bar{\sigma}}
d^{}_{\sigma}; d^\dagger_\sigma} -   J^{\ast}_{n {\bar{\sigma}}}
   \qavens{ n^{}_{d\bar{\sigma}} d^{}_\sigma ; d_{\sigma}^{\dagger} }
\nonumber  \\ & + & \sum_{m} \left [J^{\ast}_{m \bar{\sigma}}
\qavens{a^{\dagger}_{m {\bar{\sigma}}} a^{}_{n\bar{\sigma}}
d^{}_\sigma ; d^{\dagger}_\sigma} - J_{m\sigma}
\qavens{d^{\dagger}_{\bar{\sigma}} a^{}_{n\bar{\sigma}}
a^{}_{m\sigma} ; d^{\dagger}_\sigma   } \right ] \,
,\label{eq:Gamma2Exact}
\end{eqnarray}
\begin{eqnarray}
\left[ -z_2 +\varepsilon_{n
\bar{\sigma}} \right]
 \qavens{a^\dagger_{n \bar{\sigma}} d^{}_{\bar{\sigma}} d^{}_{\sigma};
 d^\dagger_\sigma} & = &
 \qav{a^\dagger_{n \bar{\sigma}} d^{}_{\bar{\sigma}} }
+ \sum_m J_{nm}^{\ast}  \qavens{a^\dagger_{m \bar{\sigma}}
d^{}_{\bar{\sigma}} d^{}_{\sigma}; d^\dagger_\sigma}
   + J_{n {\bar{\sigma}}} \qavens{ n^{}_{d\bar{\sigma}} d^{}_\sigma ;
   d_{\sigma}^{\dagger} }
\nonumber \\  &- & \sum_{m} \left [ J_{m \sigma}
\qavens{a^\dagger_{n \bar{\sigma}} d^{}_{\bar{\sigma}} a^{}_{m
\sigma}; d^{\dagger}_\sigma} - J_{m \bar{\sigma}}
\qavens{a^\dagger_{n \bar{\sigma}} d^{}_{\sigma} a^{}_{m
\bar{\sigma}}; d^{\dagger}_\sigma} \right ]\, ,
\label{eq:Gamma3Exact}
\end{eqnarray}
where we have introduced the definitions
\begin{eqnarray}
z_1 \equiv z-2 \sigma h ,\ \ \  z_2 \equiv -z +2 \epsilon_0 +U\ .
\end{eqnarray}
The EOM (\ref{eq:Gamma1Exact})--(\ref{eq:Gamma3Exact}) include
4-operator Green functions in which only two of the operators are
dot operators. Those  are decoupled as detailed in Eq.
(\ref{trunc}). One then finds
\begin{align}
-\sum_n J_{n\sigma} \qavens{a_{n\sigma} n_{\bar{\sigma}}; d^\dagger_\sigma}  = &
\Sigma_{\sigma} \qavens{ n_{\bar{\sigma}} d_\sigma ; d_{\sigma}^{\dagger} }
+ \sum_{nmm'} J_{n\sigma} \mathcal{M}_{nm\sigma}
   \qavens{a^{}_{m\sigma} ; d^{\dagger}_\sigma} \left [ J_{m'\bar{\sigma}}
   \qav{d^{\dagger}_{\bar{\sigma}} a^{}_{m' \bar{\sigma}}}
   - J^{\ast}_{m' \bar{\sigma}}
   \qav{a_{m' \bar{\sigma}}^\dagger d^{}_{\bar{\sigma}} }
   \right ]  ,
   \label{eq:Gamma1Decoup} \\
-  \sum_n J_{n\bar{\sigma}} \qavens{d^{\dagger}_{\bar{\sigma}}
a^{}_{n\bar{\sigma}} d^{}_{\sigma}; d^\dagger_\sigma}  = &
  (1+\Sigma_{\sigma} G_\sigma) P_{\bar{\sigma}}(z_1)
+  \Sigma_{\bar{\sigma}}(z_1) \qavens{ n^{}_{\bar{\sigma}}
 d^{}_\sigma ; d_{\sigma}^{\dagger} }
  -     G_{\sigma} Q_{\bar{\sigma}}(z_1)
  \, ,
  \label{eq:Gamma2Decoup} \\
  \sum_n J^{\ast}_{n\bar{\sigma}}   \qavens{a^\dagger_{n \bar{\sigma}}
  d^{}_{\bar{\sigma}} d^{}_{\sigma}; d^\dagger_\sigma}
   = &   (1+\Sigma_{\sigma} G_\sigma) P_{\bar{\sigma}}(z_2)
   - \Sigma_{\bar{\sigma}}(z_2) \qavens{ n^{}_{\bar{\sigma}} d^{}_\sigma ;
    d_{\sigma}^{\dagger} }
   +  G_{\sigma} Q_{\bar{\sigma}}(z_2) \, ,
     \label{eq:Gamma3Decoup}
\end{align}
where we have introduced
\begin{align}
  P_{\sigma}(z)& \equiv  -\sum_{nm} J_{n\sigma} \mathcal{M}_{nm\sigma}(z)
  \qav{d^\dagger_{\sigma} a{}_{m\sigma}}
 =  -\sum_{mn} \qav{a^\dagger_{m \sigma} d^{}_{\sigma} }
 \mathcal{M}_{mn\sigma}(z)
J^{\ast}_{n{\sigma}}  \, ,
  \label{eq:Pdef} \\
  Q_{\sigma}(z) & \equiv  \sum_{nmm'} J_{n\sigma}
  \mathcal{M}_{nm\sigma}(z) \qav{a^{\dagger}_{m' \sigma} a^{}_{m\sigma} }
  J^{\ast}_{m' \sigma}
   =  \sum_{nm m'}  J_{m' \sigma}
  \qav{a^\dagger_{m {\sigma}} a^{}_{m' {\sigma}}}
\mathcal{M}_{mn{\sigma}}(z) J^{\ast}_{n{\sigma}}\, .
\label{eq:Qdef}
\end{align}
The second equality in each of   Eqs.~\eqref{eq:Pdef} and
\eqref{eq:Qdef} is justified below.

Examining Eqs. ~\eqref{eq:Gamma1Decoup}, (\ref{eq:Pdef}) and
(\ref{eq:Qdef}) reveals that one needs to find thermal averages of
two types, the ones belonging to two network operators, $\qav{
a^\dagger_{m \sigma} a^{}_{m' \sigma}}$, and the ones
consisting of a dot and a network operator, $ \qav{
d^\dagger_{\sigma} a^{}_{m\sigma}}$. These are found [see
Eq. (\ref{eq:FDtheorem})] from the corresponding Green's
functions, whose EOM are  given by Eq.~\eqref{eq:cdExact} and
\begin{align}
\left( z- \varepsilon_{n\sigma} \right) \qavens{ d^{}_\sigma;
a^{\dagger}_{n \sigma} } = &
 - \sum_m  \qavens{ d^{}_\sigma; a^{\dagger}_{m \sigma} } J_{mn}
  -  J_{n\sigma} G_{\sigma} \, ,
   \label{eq:dcExact} \\
\left( z- \varepsilon_{m\sigma} \right) \qavens{ a^{}_{m \sigma} ;
a^{\dagger}_{n \sigma} } = & \, \delta_{mn} -\sum_{m'} J_{mm'}
\qavens{ a^{}_{m' \sigma} ; a^{\dagger}_{n \sigma}}-
   J^{\ast}_{m\sigma} \qavens{ d^{}_\sigma; a^{\dagger}_{n \sigma} }  \, .
   \label{eq:ccExact}
\end{align}
Their solutions in terms of the inverse matrix
$\mathcal{M}_{nm\sigma}$ [see Eq. (\ref{MMATRIX})] are
\begin{align}
\qavens{ d^{}_\sigma; a^{\dagger}_{n \sigma} } & = - \sum_m
J_{m\sigma} \mathcal{M}_{mn\sigma} G_{\sigma}
\label{eq:dcsolved} \, , \\
\qavens{ a^{}_{m \sigma} ; a^{\dagger}_{n \sigma} } & =
\mathcal{M}_{mn\sigma}+ \sum_{ln'}  \mathcal{M}_{ml\sigma}
J^{\ast}_{l\sigma} J_{n'\sigma}
 \mathcal{M}_{n'n\sigma} G_{\sigma} \, . \label{eq:ccsolved}
\end{align}
Employing these solutions to obtain the thermal averages appearing
in Eq.~\eqref{eq:Gamma1Decoup}, we find
$\sum_{m'}J_{m'\bar{\sigma}} \qav{d^{\dagger}_{\bar{\sigma}}
a^{}_{m' \bar{\sigma}}}=\sum_{m'} J^{\ast}_{m' \bar{\sigma}}
\qav{a_{m' \bar{\sigma}}^\dagger d^{}_{\bar{\sigma}} } $.
Consequently, the terms in the square brackets of
Eq.~\eqref{eq:Gamma1Decoup} are cancelled. Next, we use the first
equality in each of the definitions \eqref{eq:Pdef} and
\eqref{eq:Qdef} together with the auxiliary Green functions
\eqref{ad} and \eqref{eq:ccsolved}, to find $P$ and $Q$ in terms
the dot Green function $G$,
\begin{align}
P_{\sigma}(z) & = i \oint_C\frac{dz'}{2\pi} f(z') \,
G_{\sigma}(z') \sum_{nm} J_{n\sigma}J^{\ast}_{m\sigma}
\left [ (z-\ham_{\text{net}}^{\sigma})^{-1}
(z'-\ham_{\text{net}}^{\sigma})^{-1}\right]_{nm}  \, ,
\label{eq:Pinterm} \\
Q_{\sigma}(z) & = i \oint_C\frac{dz'}{2\pi} f(z') \,\left [ 1+
\Sigma_{\sigma}(z') G_{\sigma}(z') \right ]\sum_{nm}
J_{n\sigma}J^{\ast}_{m\sigma} \left [ (z-\ham_{\text{net}}^{\sigma})^{-1}
(z'-\ham_{\text{net}}^{\sigma})^{-1}\right]_{nm} \, ,
\label{eq:Qinterm}
\end{align}
where we have used Eq.~\eqref{MMATRIX}. Since
\begin{eqnarray}
(z-\ham_{\text{net}}^{\sigma})^{-1}(z-\ham_{\text{net}}^{\sigma})^{-1}=
\frac{(z'-\ham_{\text{net}}^{\sigma})^{-1}-
(z-\ham_{\text{net}}^{\sigma})^{-1}}{z-z'}\ ,
\end{eqnarray}
we can use Eq. (\ref{Sigmadef}) for the non-interacting
self-energy, to write the functions $P$ and $Q$ in terms of that
self-energy,
\begin{eqnarray}
P_{\sigma}(z)&=& \lim_{\eta'\to 0} \frac{i}{2 \pi} \int d \omega
\frac{f(\omega ) }{z-\omega }\Bigl (G_{\sigma}(\omega+i \eta')
\left [\Sigma_{\sigma}(\omega+i \eta')-\Sigma_{\sigma}(z) \right]
- G_{\sigma}(\omega-i \eta') \left [\Sigma_{\sigma}(\omega-i
\eta')-\Sigma_{\sigma}(z) \right ]\Bigr )
\nonumber\\
&&\equiv \frac{i}{2 \pi}\oint_C f(w) G_{\sigma}(w)
\frac{\Sigma_{\sigma}(w)-\Sigma_{\sigma}(z)}{z-w} d w\, ,
\nonumber\\
Q_{\sigma}(z)&=& \lim_{\eta'\to 0} \frac{i}{2 \pi} \int d \omega
\frac{f(\omega )}{z-\omega}\nonumber\\
\times\Bigl ( [1&+&\Sigma_{\sigma}(\omega+i \eta')
G_{\sigma}(\omega+i \eta')] \left [\Sigma_{\sigma}(\omega+i
\eta')-\Sigma_{\sigma}(z) \right] - [1+\Sigma_{\sigma}(\omega-i
\eta')G_{\sigma}(\omega-i \eta')] \left [\Sigma_{\sigma}(\omega-i
\eta')-\Sigma_{\sigma}(z) \right ]\Bigr )
\nonumber\\
&&\equiv \frac{i}{2 \pi}\oint_C
f(w)[1+\Sigma_{\sigma}(w)G_{\sigma}(w)]
\frac{\Sigma_{\sigma}(w)-\Sigma_{\sigma}(z)}{z-w} d w\,  .
\label{EXPLICIT}
\end{eqnarray}
\end{widetext}
Here and elsewhere the imaginary part of $z$ is always greater
than $\eta'$, so that the contour $C$ never encircles the pole at
$w =z$. Note that  using the same procedure employing the second
equalities in Eqs.~\eqref{eq:Pdef} and \eqref{eq:Qdef} gives again
Eqs.~\eqref{eq:Pinterm} and \eqref{eq:Qinterm}, thus proving that
the two definitions of $P$ and $Q$ in Eqs.~\eqref{eq:Pdef} and
\eqref{eq:Qdef} are equivalent. Equation ~\eqref{GammadExact} can
be now easily solved. Inserting the solution into
Eq.~\eqref{GdExact} leads to the expression for the dot Green
function,  Eq.~\eqref{eq:GfiniteU}.

In treating the functions $P$ and $Q$ in Sec. \ref{sec:phsym} we
have employed several properties of the complex integrals
appearing in Eqs. (\ref{EXPLICIT}). Consider for example the
integral
\begin{align}
\frac{i}{2 \pi}\oint_C G(w) \frac{\Sigma(w)-\Sigma(z)}{z-w} d w\ .
\label{eq:integralexample}
\end{align}
Since $G(w)$ and $\Sigma (w)$ have no singularities except for a
cut along the real axis, it is expedient to complete each half of
the contour $C$ by a large-radius semi-circle in the upper and
lower half-planes.  Then, by the residue theorem, Eq.
\eqref{eq:integralexample} vanishes provided that $G(w)$ falls as
$w^{-1}$ or faster at $w \to \infty$. This means that the Fermi
function $f(w)$ in the definition of  $P$ can be replaced by
$f(w)+\text{const}$. Another important case is when $G(w)$ is
replaced by 1 in Eq. \eqref{eq:integralexample}. In this case the
contribution of the  semi-circles does not vanish, and the
integral \eqref{eq:integralexample} gives  $\Sigma(z)$.

An alternative to the fully self-consistent treatment investigated
in this paper has been proposed in Ref.~~\onlinecite{Meir91ANDGF}.
There, the averages of the form $ \qav{d^\dagger_{\sigma}
a^{}_{m\sigma}}$ were ignored, and those of the type
$\qav{a^\dagger_{m \sigma} a^{}_{m' \sigma}}$ were approximated by
$\delta_{mm'}\qav{ a^\dagger_{m \sigma} a^{}_{m \sigma}}$. Namely,
$\ham_{\text{net-dot}}$ was put to zero in the calculation of the
averages. Upon such an approximation, $P_{\sigma}(z) \simeq 0$,
and
\begin{align}
 Q_{\sigma}(z) \simeq
 \int \im \Bigl [\frac{\Sigma_{\sigma}(\omega-i\eta)}{\pi}\Bigr ]
 \frac{f(\omega)}{z-\omega} \, \, d \omega \ ,
\end{align}
reproducing Eq.~(8) of Ref.~~\onlinecite{Meir91ANDGF}. Another
approximation of Eqs. (\ref{eq:Pinterm}) and (\ref{eq:Qinterm}),
originally due to Lacroix\cite{Lacroix81}, has been recently
analyzed in the context of quantum dots in
Ref.~~\onlinecite{EntinAharonyMeir04}. In this approximation one
assumes the dot Green function to vary smoothly enough over the
integration regimes in Eqs. (\ref{eq:Pinterm}) and
(\ref{eq:Qinterm}), so that it can be taken out of the integrals.
This \emph{ansatz} reduces Eq.~\eqref{eq:GfiniteU} of the text to
a quadratic form. These two approximate solutions are discussed in
Secs.~\ref{sec:finiteU} and \ref{sec:Uinf}.

\begin{widetext}
\section{Derivation of the polynomial function $R$}
\label{ASEX}
As explained in Sec. \ref{sec:Uinf}, the function
\begin{eqnarray}
R(z)&\equiv&\Phi_{\sigma} (z)\widetilde{\Phi}_{\sigma}(z)-{\rm
X}_{\sigma}(z)\widetilde{\rm
X}_{\sigma}(z)\nonumber\\
&=&i(\Gamma_{\sigma}+\Gamma_{\bs})\Bigl [
z-\epsilon_{0}-i\Gamma_{\bs }-I_{\sigma}(z+\sigma h)\Bigr ]\Bigl
[\qav{ 1-n_{d\bs}} -\frac{\Gamma_{\bs}}{\pi}\int d\omega
\frac{f(\omega )G^{-}_{\bs}(\omega )}{z+\bs h-\omega}\Bigr
]\nonumber\\
&-&i(\Gamma_{\sigma}+\Gamma_{\bs})\Bigl [
z-\epsilon_{0}+i\Gamma_{\sigma }-I_{\bs}(z+\bs h)\Bigr ]\Bigl
[\qav{1-n_{d\sigma}} -\frac{\Gamma_{\sigma}}{\pi}\int
d\omega \frac{f(\omega )G^{+}_{\sigma}(\omega )}{z+\sigma
h-\omega}\Bigr
]\nonumber\\
&-&\bigl ( \Gamma_{\sigma}+\Gamma_{\bs}\bigr )^{2}\Bigl [\qav{
1-n_{d\bs}} \qav{1-n_{d\sigma}}
-\frac{\Gamma_{\bs}}{\pi}\int d\omega \frac{f(\omega
)G^{-}_{\bs}(\omega )}{z+\bs
h-\omega}\times\frac{\Gamma_{\sigma}}{\pi}\int d\omega
\frac{f(\omega )G^{+}_{\sigma}(\omega )}{z+\sigma h-\omega}\Bigr ]
\label{R}
\end{eqnarray}
is non-singular across the real axis, and its only singular point
is at $z=\infty$. This observation enables one to solve for the
Green function in terms of the functions $\Phi$ and
$\widetilde{\Phi}$. Here we examine $R(z)$ in some detail, and
also derive the first two terms of its polynomial expansion.

In the limit $z\rightarrow\infty$, the function $I_{\sigma}(z)$,
Eq. (\ref{eq:Iin}), is given by
\begin{eqnarray}
I_{\sigma}(z)\sim\frac{\Gamma_{\sigma}}{\pi}\ln\frac{D_{\sigma}}{z}\,
.\label{IAPP}
\end{eqnarray}
The $z\rightarrow\infty$ limit of the integrals appearing in Eq.
(\ref{R}) has to be taken with care. We write the Green functions
appearing in the integrands there in the form
$G^{\pm}_{\sigma}(\omega )=\re G^{+}_{\sigma}(\omega )\pm
i \im G^{+}_{\sigma}(\omega )$, and use $\qav{
n_{d\sigma}}=-(1/\pi )\int d\omega f(\omega ) \im
G^{+}_{\sigma}(\omega )$ to obtain
\begin{eqnarray}
\frac{\Gamma_{\sigma}}{\pi}\int d\omega\frac{f(\omega
)G^{\pm}_{\sigma}(\omega )}{z-\omega }\sim\mp
i\frac{\Gamma_{\sigma}\qav{
n_{d\sigma}}}{z}+A_{\sigma}(z)\, ,\label{FAPP}
\end{eqnarray}
where
\begin{eqnarray}
A_{\sigma}(z)=\frac{\Gamma_{\sigma}}{\pi}\int d\omega
\frac{f(\omega ) \re G^{+}_{\sigma}(\omega )}{z-\omega}\, .
\label{AA}
\end{eqnarray}
Inserting Eqs. (\ref{IAPP}) and (\ref{FAPP}) into Eq. (\ref{R}),
the terms which survive the $z\rightarrow\infty$ limit are
\begin{eqnarray}
R(z)&\sim &(\Gamma_{\sigma}+\Gamma_{\bs})^{2}\bigl (\qav{
n_{d\sigma}} +\qav{n_{d\bs}} -\qav{
n_{d\sigma}} \qav{n_{d\bs}}\bigr
)+i(\Gamma_{\sigma}+\Gamma_{\bs})\bigl (z-\epsilon_{0}\bigr
)\qav{n_{d\sigma}-n_{d\bs}}\nonumber\\
&+&i(\Gamma_{\sigma}+\Gamma_{\bs})\Bigl (z\bigl
(A_{\sigma}(z)-A_{\bs}(z)\bigr )-\qav{
1-n_{d\bs}}\frac{\Gamma_{\sigma}}{\pi}\ln
\frac{D_{\sigma}}{z}+\qav{
1-n_{d\sigma}} \frac{\Gamma_{\bs}}{\pi}\ln
\frac{D_{\bs}}{z}\Bigr )\, .
\end{eqnarray}
According to the discussion in Sec. \ref{sec:Uinf}, the terms
logarithmic in $z$ have to disappear. The integral giving
$A_{\sigma}(z)$, Eq. (\ref{AA}), is well behaved on the positive
$\omega$ axis, since then for large $\omega$ the Fermi function
makes it  convergent. For very large negative $\omega $ values,
$\re G^{+}_{\sigma}(\omega )\rightarrow\qav{
1-n_{d\bs}}/\omega$, and as a result, the contribution from
that part of the integration to $A_{\sigma}(z)$ is $\qav{
1-n_{d\bs}}(\Gamma_{\sigma}/\pi )(1/z) \ln (\zeta_{\sigma}
/z)$, where $\zeta_{\sigma}\lesssim D_{\sigma}$. Hence, the terms
logarithmic in $z$ are cancelled. In our calculations, we have
used
\begin{eqnarray}\label{eq:Afinal}
A_{\sigma}(z)\sim -\frac{\Gamma_{\sigma}}{\pi}\qav{
1-n_{d\bs}}\ln \frac{z}{D_{\sigma}} +\frac{b_{\sigma}}{z} \,
,\label{ASIG}
\end{eqnarray}
and have determined the coefficient $b_{\sigma}$ self-consistently
(see next Appendix). In this way we find
\begin{eqnarray}
R(z)&= &(\Gamma_{\sigma}+\Gamma_{\bs})^{2}\Bigl [\qav{
n_{d\sigma}} +\qav{n_{d\bs}} - \qav{
n_{d\sigma}} \qav{n_{d\bs}} \Bigr
]+i(\Gamma_{\sigma}+\Gamma_{\bs})\Bigl [\bigl (z-\epsilon_{0}\bigr
)\qav{n_{d\sigma}-n_{d\bs}}+b_{\sigma}-b_{\bs}\Bigr ]\,
.\label{RFIN}
\end{eqnarray}

\section{Details of the exact solution}
\label{RH} This Appendix is devoted to the analysis of the exact
solution for the self-consistent EOM,  and in particular to the
determination of the unknown coefficients $b_{\sigma}$ [see Eqs.
(\ref{ASIG}) and (\ref{RFIN})], $a$ and $\widetilde{a}$ [see Eqs.
(\ref{SOL})]. This is accomplished by expanding the solution at
large frequencies, and equating the coefficients with those of the
desired functions $\Phi$ and $\widetilde{\Phi}$. We give the
details for $\Phi$; those of the `tilde' solution are obtained
analogously.

As has been the case for the integral \eqref{FAPP}, the large
negative part of the integral defining $M_{\sigma}(z)$ has to be
taken with care. To this end we write
\begin{align}\label{eq:Msplit}
 M_{\sigma}(z) \equiv \int \Bigl (-\frac{d \omega}{2 \pi i}\Bigr )
 \frac{\ln H_{\sigma}(\omega) }{z-\omega} =
\int_{-\infty}^{+\infty} \Bigl (-\frac{d \omega}{2 \pi i}\Bigr )
\frac{\ln H_{\sigma}(\omega) - \Theta(-D_{\bs} - \omega)
F_{\sigma}(\omega) (-2 \pi i)}{z-\omega} +
\int_{-\infty}^{-D_{\bs}} \frac{ F_{\sigma}(\omega)\, d
\omega}{z-\omega} \, ,
\end{align}
where the function $F_{\sigma}(\omega)$ is defined in such a way
that the use of the geometric series $1/(z-\omega)=z^{-1}+z^{-2}
\omega +\mathcal{O}(z^{-3})$ in the first integral of Eq.~\eqref{eq:Msplit}
results in convergent integrals. It is sufficient to include in
$F_{\sigma}(\omega)$ the most slowly-decaying terms of $\ln
H(\omega)/(-2\pi i )$, which are obtained by expanding $H(\omega)$
for large negative $\omega$,
\begin{align}
F_{\sigma}(\omega) & = -\frac{\Gamma_{\bar{\sigma}}}{\pi \omega }
-\frac{\Gamma_{\bar{\sigma}}}{\pi \omega^2 }  \left [
\frac{\Gamma_{\bs}}{\pi} \ln \frac{D_{\bs}}{|\omega|} +\epsilon_0
 - i (\Gamma_{\sigma} + \Gamma_{\bs}) \qav{n_{d \sigma}} + i \Gamma_{\bs}
\right ] \, .\label{FEQ}
\end{align}
[In this expansion one has to include terms of the order
$\mathcal{O}(\omega^{-2})$ because of the linear term in the prefactor in
Eq. \eqref{SOL}.] Using Eq. (\ref{FEQ}), the second integral in
Eq.~\eqref{eq:Msplit} is obtained explicitly, and then expanded up
to order $z^{-2}$,
\begin{align}\label{eq:FintegralResult}
  \int_{-\infty}^{-D_{\bs}} \frac{ F_{\sigma}(\omega)\, d \omega}{z-\omega}
\sim & \frac{\Gamma_{\bs}}{\pi z} \ln \frac{z}{D_{\bs}} - \frac{1}{2}
\left ( \frac{\Gamma_{\bs}}{\pi z} \ln \frac{z}{D_{\bs}}
\right )^2 + \frac{\Gamma_{\bs}^2}{\pi^2 \, D_{\bs} \, z} -
\frac{\Gamma_{\bs}^2}{ 6\, z^2}
\nonumber \\
    & +\frac{D_{\bs} \Gamma_{\bs}}{ \pi \, z^2}  -\frac{\Gamma_{\bs}}{\pi} \left [
\epsilon_0
 - i (\Gamma_{\sigma} + \Gamma_{\bs}) \qav{n_{d \sigma}} + i \Gamma_{\bs}
\right ] \left [ \frac{1}{z\, D_{\bs}} - \frac{1}{z^2} \ln \frac{z}{D_{\bs}} \right]
\, .
\end{align}
As a result, the asymptotic expansion of the function
$M_{\sigma}(z)$ becomes
\begin{align}
  M_{\sigma}(z)
 \sim  & \bigl [ \alpha_{\sigma} + (\Gamma_{\bs}/\pi) \, \ln (z/D_{\bs}) \bigr ]/z
   + \nonumber \\
& \bigl \{ \beta_{\sigma}  + (\Gamma_{\bs}/\pi) \,
\bigl [ \epsilon_0
 - i (\Gamma_{\sigma} + \Gamma_{\bs}) \qav{n_{d \sigma}} + i \Gamma_{\bs}
\bigr ] \,
  \ln (z/D_{\bs})
    -  (\Gamma_{\bs}/\pi)^2 (1/2) \ln^2 (z/D_{\bs})
  \bigr \}/z^2 \label{eq:Mexpanded} \, ,
\end{align}
where the coefficients $\alpha_{\sigma}$ and $\beta_{\sigma}$ are
given by
\begin{align}
  \alpha_{\sigma}&  = \frac{\Gamma_{\bs}^2}{\pi^2 D_{\bs}} - \frac{\Gamma_{\bs}}{\pi D_{\bs}}
\bigl [ \epsilon_0
 - i (\Gamma_{\sigma} + \Gamma_{\bs}) \qav{n_{d \sigma}} + i \Gamma_{\bs}
\bigr ]+
\int_{-\infty}^{+\infty} \!\!\!
\left [ i \ln H_{\sigma}(\omega) /( 2\pi)
- \Theta(-D_{\bs} - \omega) F_{\sigma}(\omega)
\right ] d \omega \, , \\
\beta_{\sigma} & =
 -\frac{\Gamma_{\bs}^2}{6}+ \frac{\Gamma_{\bs}  D_{\bs}}{\pi} +
\int_{-\infty}^{+\infty} \!\!\!
\omega \left [ i \ln H_{\sigma}(\omega) /( 2\pi)
- \Theta(-D_{\bs} - \omega) F_{\sigma}(\omega)
\right ] d \omega \, .
\end{align}
An analogous calculation gives $\widetilde{F}_{\sigma}(\omega)
=F_{\bar{\sigma}}(\omega)^{\ast}$, leading to
\begin{align}
    \widetilde{M}_{\sigma}(z)
 \sim  & \bigl [ \widetilde{\alpha}_{\sigma} + (\Gamma_{\sigma}/\pi) \, \ln (z/D_{\sigma}) \bigr ]/z
   + \nonumber \\
& \bigl \{ \widetilde{\beta}_{\sigma}  + (\Gamma_{\sigma}/\pi) \,
\bigl [ \epsilon_0
 + i (\Gamma_{\sigma} + \Gamma_{\bs}) \qav{n_{d \bs}} - i \Gamma_{\sigma}
\bigr ] \,
  \ln (z/D_{\sigma})
    -  (\Gamma_{\sigma}/\pi)^2 (1/2) \ln^2 (z/D_{\sigma})
  \bigr \}/z^2 \label{eq:MTildeexpanded} \, ,
\\
\widetilde{\alpha}_{\sigma}  = & \frac{\Gamma_{\sigma}^2}{\pi^2 D_{\sigma}} - \frac{\Gamma_{\sigma}}{\pi D_{\sigma}}
\bigl [ \epsilon_0
 + i (\Gamma_{\sigma} + \Gamma_{\bs}) \qav{n_{d \bs}} - i \Gamma_{\sigma}
\bigr ]+
\int_{-\infty}^{+\infty} \!
\left [ i \ln \widetilde{H}_{\sigma}(\omega) /( 2\pi)
- \Theta(-D_{\sigma} - \omega) \widetilde{F}_{\sigma}(\omega)
\right ] d \omega \, , \\
\widetilde{\beta}_{\sigma}  = &
 -\frac{\Gamma_{\sigma}^2}{6}+ \frac{\Gamma_{\sigma}  D_{\sigma}}{\pi} +
\int_{-\infty}^{+\infty} \!\!\!
\omega \left [ i \ln \widetilde{H}_{\sigma}(\omega) /( 2\pi)
- \Theta(-D_{\sigma} - \omega) \widetilde{F}_{\sigma}(\omega)
\right ] d \omega \, .
\end{align}

Finally, we use Eqs.~\eqref{eq:Mexpanded} and
\eqref{eq:MTildeexpanded} in Eq.~\eqref{SOL}, and then compare
term by term with the expansion of Eq.~\eqref{PSI}. This procedure
determines the coefficients $a$ and $\widetilde{a}$,
\begin{align}
  a & = \alpha_{\sigma}+\epsilon_0-i \Gamma_{\sigma} \, , \quad
  \widetilde{a} = \widetilde{\alpha}_{\sigma}+\epsilon_0 +i \Gamma_{\bs} \, ,
\label{eq:a12}
\end{align}
and gives  the  self-consistency equations
\begin{align}
  b_{\bar{\sigma}} + i \Gamma_{\bs}\qav{n_{d\bs}} & = i
\left (\Gamma_{\sigma} + \Gamma_{\bs} \right)^{-1}
\bigl[ {\beta}_{\sigma}-a \, {\alpha}_{\sigma} + {\alpha}_{\sigma}^2/2
  + \Gamma_{\bs}  h /({2 \pi}) \bigr ] \label{eq:bdown}  \, , \\
b_{\sigma} - i \Gamma_{\sigma }\qav{n_{d\sigma}}
& = -i \left (\Gamma_{\sigma} + \Gamma_{\bs} \right)^{-1}
\bigl[ \widetilde{\beta}_{\sigma}-\widetilde{a} \, \widetilde{\alpha}_{\sigma} +
\widetilde{\alpha}_{\sigma}^2/2
  - \Gamma_{\sigma}  h /({2 \pi}) \bigr ]  \, . \label{eq:bup}
\end{align}
\end{widetext}
In the case of  a full spin-symmetry (including $h=0$), one has
$\widetilde{\alpha}=\alpha^{\ast}$, $\widetilde{\beta}=
\beta^{\ast}$, and $\widetilde{a}=a^{\ast}$, and then
Eq.~\eqref{eq:bdown} and Eq.~\eqref{eq:bup} become complex
conjugate.

For a given set of parameters, Eqs.~\eqref{eq:bdown} and
\eqref{eq:bup} are solved numerically for $b_{\sigma}$  and
$\qav{n_{d\sigma}}$ by an iterative Newton-Raphson algorithm. The
initial values are chosen from the solution of the non-interacting
($U=0$) problem.

% Bibliography has been prepared with the help of BibTeX


\begin{thebibliography}{35}
\bibitem{foot2}
Keeping the EOM of these Green functions and decoupling the
resulting higher-order EOM has been proposed in
Ref.~~\onlinecite{Luo99} as a possible way to improve the present
approximation. We have tried to investigate this proposal but
could not reproduce  the results of that paper. Instead, we have
found that the scheme becomes un-tractable.
\bibitem{footaa}  In fact, in order to obtain the result (\ref{eq:Iin})
it suffices to make the assumption (\ref{ASSU}) only over an
energy scale which is much smaller than the bandwidth, say
$[-D_\sigma^{\text{low}}; D_\sigma^{\text{up}} ]$. Possible energy
dependencies of the self-energy on larger scales can then be
incorporated into $D_{\sigma}$, by normalizing the lower cutoff on
the band, $D_{\sigma} \equiv D_\sigma^{\text{low}} \exp
\int_{-D_\sigma^{\text{low}}}^0 [1-\im \Sigma_{\sigma}^{-}
(\omega)/\Gamma_{\sigma}]/\omega \, d\omega$. This may be
necessary when the dot is embedded in a complex mesoscopic
network.
\bibitem{foot3}
In such a case the so-called index of the Riemann-Hilbert problem
\cite{Muskhlishvili53} is zero. If the integration limits in
Eq.~\eqref{SOL} are to be kept finite, one has to assume $D\equiv
D_\sigma$ and possibly consider additional multiplicative powers
of $(z \pm D)$, see Ref.~~\onlinecite{BloomfieldHamann67}. We
ignore these options since they only affect the corrections of
order $\omega/D_\sigma$.
\end{thebibliography}
\end{document}